\documentclass[12pt,reqno]{article}
\usepackage[english]{babel}
\usepackage{amsmath,amsthm,commath,mathrsfs,amssymb,extarrows}
\usepackage{dsfont}
\usepackage{relsize}

\usepackage{tikz}
\usetikzlibrary{arrows,intersections}

\usepackage[colorlinks = true,urlcolor = blue,citecolor = blue,breaklinks]{hyperref}

\usepackage{amsfonts}
\usepackage{graphicx}
\usepackage{enumerate}
\usepackage{subcaption}
\usepackage[right,pagewise,displaymath, mathlines]{lineno}
\usepackage{epstopdf}
\usepackage{color}
\usepackage{multirow}
\usepackage{authblk}
\usepackage[round]{natbib}
\usepackage{float}
\restylefloat{figure,table}

\usepackage{booktabs}

\usepackage{colortbl}

\usepackage{geometry}
\numberwithin{equation}{section}
\numberwithin{figure}{section}
\numberwithin{table}{section}

\newtheorem{theorem}{Theorem}[section]

\theoremstyle{definition}

\newtheorem{example}{Example}[section]
\newtheorem{note}{Note}[section]

\numberwithin{equation}{section}

\definecolor{darkread}{rgb}{0.7, 0, 0}
\definecolor{darkbrown}{rgb}{0.55, 0.2, 0.15}
\definecolor{darkblue}{rgb}{0.1,0.1,0.6}
\definecolor{darkgreen}{rgb}{0.1,0.5,0.2}

\DeclareMathOperator*{\argmax}{arg\,max}

\setcitestyle{square,numbers}
\makeatletter
\renewcommand\@biblabel[1]{#1.}
\makeatother

\begin{document}

\begin{center}

{\Large\bf Tail maximal dependence in bivariate models: estimation and applications}

\end{center}

\begin{center}

Ning Sun$^{a}$,
Chen Yang$^{b,*}$, and
Ri\v cardas Zitikis$^{a,c}$

\end{center}

\begin{center}\it

$^{a}$School of Mathematical and Statistical Sciences, Western University, London, Ontario~N6A~5B7, Canada

$^{b}$Department of Population Health Science and Policy, Icahn School of Medicine at Mount Sinai, New York, NY 10029, U.S.A.

$^{c}$Risk and Insurance Studies Centre, York University,  Toronto, Ontario M3J 1P3, Canada

$^{*}$e-mail: Chen.Yang@mountsinai.org

\end{center}

\begin{center}\sf

To appear in \textit{Mathematical Methods of Statistics}


\end{center}

\begin{quote}
{\bf Abstract.}
Assessing dependence within co-movements of financial instruments has been of much interest in risk management. Typically, indices of tail dependence are used to quantify the strength of such dependence, although many of them underestimate the strength. Hence, we advocate the use of indices of maximal tail dependence, and for this reason we also develop a statistical procedure for estimating the indices. We illustrate the procedure using simulated and real data sets.
\medskip

\end{quote}

\section{Introduction}

The phenomenon of extreme co-movements manifests in a variety of problems~\citep{chbs2004}.  In financial management, this phenomenon arises when, for example, dealing with contagion \citep{Pericoli2003} and developing risk mitigation strategies such as portfolio diversification \citep{Durante2014}. These and other applications have inspired prolific studies of how to assess dependence within extreme co-movements~\citep{BJW2015, WKM2015}. 

For this purpose, researchers have often employed tail dependence indices \citep{chbs2004,Frahm2005,CLV2013}. Their definitions rely on the behaviour of the copula $C:[0,1]\times [0,1]\to[0,1]$, which arises from paired financial instruments,  along the diagonal path $(u,u)_{0\le u \le 1}$ near the vertex $\mathbf{0}:=(0,0)$ of the unit square $[0,1]\times [0,1]$. In particular, the parameter $\kappa\in[1,\infty)$ in the following equation (assuming that it holds)
\begin{equation}\label{tail eq}
C(u,u) = \ell(u)u^{\kappa}
\end{equation}
is called the lower tail order, henceforth called the tail order of \textit{diagonal} dependence (TODD), where $\ell$ is a slowly varying at $0$ function. The modifier \textit{diagonal} has been added to emphasize the equality of the two copula arguments on the left-hand side of equation~\eqref{tail eq}. Recall also that $\ell$ is called slowly varying at $0$ whenever the statement
\begin{equation}
{\ell(qx)\over \ell(q)}\to 1 \quad\text{when} \quad q\downarrow 0,
\label{ratio-ells}
\end{equation}
holds irrespective of $x>0$.

Interestingly, the asymptotic behaviour of $C(u,u)$  may or may not reflect the \textit{maximal} strength of tail dependence, as has been pointed out and illustrated by \citet{Furman2015,FKSZ2016}. Hence, it becomes natural to seek a path, called  path of maximal tail dependence (MTD), along which the copula $C$ behaves like $\ell^*(u)u^{\kappa^*}$ for the smallest possible $\kappa^*$ and some slowly varying at $0$ function $\ell^*$.
We note that this MTD path may not be unique, and it may not coincide with the diagonal path $(u,u)_{0\le u \le 1}$. Henceforth, we call $\kappa^*$ the tail order of \textit{maximal} dependence (TOMD), whose rigorous description will be given in the next section.

For the TOMD $\kappa^*$, basic statistical inference has been developed by \citet{SYZ2020} in the case of independent and identically distributed (iid) paired data. In practice, however, co-movements often arise from time-indexed stochastic processes, such as time series
\citep{WKM2015},
whose generated data are inherently dependent.
The iid-based results, therefore, need to be adjusted and extended. Fortunately, the block structure of the estimators that we have developed facilitates the task, due to the fact that practically relevant time series give rise to nearly-independent data blocks when the gaps between the blocks are sufficiently wide. Nevertheless, despite this simple explanation of the idea, serious technical work remains to be done, which  we accomplish in the present paper. The very definitions of the proposed estimators, as well as the intuition behind them that we shall describe in following sections, will help to clearly see the encountered complexities and ways to tackle them.

\begin{note}
To alleviate some of the challenges, researchers have recently sought ways for employing tangent lines of the MTD paths, and then estimating the slopes of those lines as proxies to the TOMD. For an enlightening reading on the topic, we refer to \citet{KKH2022}.
\end{note}

The rest of this paper is organized as follows. In Section~\ref{new-estimator} we introduce TOMD and TODD estimators.
In Section~\ref{application} we discuss the estimators and prepare them for the analysis of real time-series data.
In Section~\ref{appl-simulations} we develop a dependent-data generating process and then illustrate and assess the performance of the estimators.
In Section~\ref{appl-0} we explore the strength of extreme co-movements of a number of financial instruments, such as exchange rates of several currencies, stock market indices,
and mixed financial instruments.  Section~\ref{sec: conclusion} concludes the paper with a summary. To facilitate readability of the paper, tedious proofs, extensive  statistical tests, tables and graphs are relegated to Appendix~\ref{append-00}.


\section{MTD paths, their tail orders, and estimators}
\label{new-estimator}

The notion of path of maximal tail dependence (MTD) is pivotal. Although it has been around for nearly a decade, we feel that we still need to introduce it in an intuitive and gradual manner, which will in turn help us to facilitate the introduction of rather complex (and inevitably so) empirical estimators of the TOMD and TODD.

\subsection{MTD paths}

We start at what we think would be the most intuitive level. Namely, given two random variables $U$ and $V$ with uniform on the unit interval $[0,1]$ marginal distributions, their copula $C(s,t)$ is the proportion (a.k.a.\, probability) of the realizations placed by the random pair $(U,V)$ into the rectangle $[0,s]\times [0,t]$, whose left-bottom vertex is the origin $(0,0)$ and the right-top vertex is the point $(s,t)$. While classical indices of tail dependence count the proportions of those $(U,V)$ realizations that are in the square $[0,u]\times [0,u]$, with $u$ near $0$, we argue that using rectangles $[0,s]\times [0,t]$ with the largest proportion of the $(U,V)$ realizations would reveal the maximal strength of tail dependence between $U$ and $V$.

Of course, we realize that the maximal proportion would be achieved when $s=1=t$, in which case the rectangle $[0,s]\times [0,t]$ turns into the unit square $[0,1]\times [0,1]$. Hence, the proper way to search for those rectangles $[0,s]\times [0,t]$ that contain the maximal proportion of the $(U,V)$ realizations is to equate their areas to that of the square $[0,u]\times [0,u]$. In this way we arrive at the equation $st=u^2$, and it is under this equation that we search for those rectangles $[0,s]\times [0,t]$ that contain the maximal proportion of the $(U,V)$ realizations. We note at this point that the square $[0,u]\times [0,u]$ may or may not have the maximal proportion of the $(U,V)$ realizations, and this is the reason why the classical measures of tail dependence, which are based on the square
$[0,u]\times [0,u]$, may or may not provide reliable information about the maximal strength of dependence between $U$ and $V$. In view of the equation $st=u^2$, the right-top vertices of the rectangles $[0,s]\times [0,t]$ give rise to curves parametrized by $u\in [0,1]$, and these curves are the MTD paths.

Still intuitively but at a higher level of mathematical rigour, we note that the role of MTD path, written generally as $(\varphi^*(u),\psi^*(u))_{0\le u\le 1}$, is for each $u\in (0,1]$ to provide a rectangle
\[
\mathcal{R}_u(\mathbf{0}):=[0,\varphi^*(u)]\times [0,\psi^*(u) ]
\]
with the following properties:
\begin{enumerate}[(1)]
\item
it is a subset of the unit square $[0,1]\times [0,1]$;
\item
has the same area as the square $[0,u]\times [0,u]$, which necessarily implies the equation $\psi^*(u) =u^2/\varphi^*(u)$;
\item
when a large number of pairs are simulated from the copula $C$, the rectangle $\mathcal{R}_u(\mathbf{0})$ contains at least as many simulated pairs as any other rectangle in the first quadrant of the plane with one of its vertices being $\mathbf{0}$ and the area equal to $u^2$.
\end{enumerate}

Now in full mathematical rigour as in \citet{Furman2015}, for any bivariate copula $C:[0,1]\times [0,1] \to [0,1]$, a path of (not necessarily maximal) tail dependence is $(\varphi(u),u^2/\varphi(u))_{0\le u\le 1}$ for a function  $\varphi:[0,1] \to [0,1]$ that satisfies the following two admissibility conditions:
\begin{enumerate}[(1)]
\item
$\varphi(u)\in[u^2, 1]$ for every $u\in[0,1]$,
\item
$\varphi(u)\to 0$ and $u^2/\varphi(u)\to 0$ when $u\downarrow 0$.
\end{enumerate}
Denote the set of all admissible functions by $\mathcal{A}$. An admissible function $\varphi^*\in \mathcal{A}$ that maximizes the functional
$\varphi \mapsto C(\varphi(u), u^2/\varphi(u))$ gives rise to a path
$(\varphi^*(u), u^2/\varphi^*(u))_{0\le u\le 1}$, which is a path of \textit{maximal} tail dependence (MTD). Since there can be several maximizing functions $\varphi^*$, there can be several MTD paths.

For any of these MTD paths, we define $\Pi^*:[0,1] \to [0,1]$ by
\[
\Pi^*(u) = C(\varphi^*(u), u^2/\varphi^*(u)).
\]
We now make the assumption that $\Pi^*(u)$ is a regularly varying function at $0$ \citep{bgt1987}. That is, there exist a parameter $\kappa^* $ and a slowly varying at $0$ function $\ell^*$ such that
\begin{equation}\label{max-tail eq}
\Pi^*(u) = \ell^*(u)u^{\kappa^*}
\end{equation}
when $u\downarrow 0$. The parameter $\kappa^*$ is called the lower tail order of \textit{maximal} dependence (TOMD).

The parameter $\kappa^*$ is unique for any given copula $C$, irrespective of the fact that there can be several admissible functions $\varphi^*$ leading to it. As illustrated by \citet{Furman2015}, the TOMD $\kappa^*$ may or may not coincide with the TODD  $\kappa$, and the Gaussian copula provides one of those rarest examples when $\kappa^*=\kappa$ \citep{FKSZ2016}.

\subsection{Tail-order estimators}

We are now in the position to introduce an empirical estimator of the TOMD $\kappa^*$. Suppose that the underlying model is a bi-variate stationary time series $(X_i,Y_i)$, $i\in \mathbb{Z}$, and let the pairs $(X_1,Y_1), \dots, (X_n,Y_n)$ be observable. (Note that we are already deviating from the framework of \citet{SYZ2020}, which assumes iid pairs.)  Hence, our data are $(x_1,y_1), \dots, (x_n,y_n)\in \mathbb{R}^2$.

Next we separate the marginal distributions from the dependence structure, about which we learn from  the bivariate pseudo-observations $(u_1,v_1), \dots, (u_n,v_n)\in [0,1]\times [0,1]$ generated by the copula $C$. These pseudo-observations give rise to a scatterplot in the unit square $[0,1]\times [0,1]$, and they also define the empirical copula
\[
C_n(u,v)={1\over n}\sum_{i=1}^n\mathds{1}\big\{u_i\le u,v_i\le v\}, \quad u,v\in [0,1].
\]

In order to mitigate the potential influence of $\ell^*$ (equation~\eqref{max-tail eq}) on the TOMD estimator, we focus on those observed pairs that fall into the rectangle
\[
\mathcal{R}_{q,n}(\mathbf{0})= [0,\varphi_n^*(q)]\times [0,q^2/\varphi_n^*(q)],
\]
where
\[
\varphi_n^*(q)=\argmax_{x\in[q^2,1]} C_n(x,q^2/x) .
\]
Choices of $q\in (0,1]$ will be discussed later in this paper for simulated as well as for real data.

\begin{note}
When $q=1$, the rectangle $\mathcal{R}_{q,n}(\mathbf{0})$ is the unit square  $[0,1]\times [0,1]$, which we used in \citet{SYZ2020} due to the absence of the slowly varying function $\ell^*$ and thus having no need for introducing a threshold to mitigate the influence of $\ell^*$ on the estimation procedure.
\end{note}

Hence, from all the pseudo observations $(u_1,v_1), \dots, (u_n,v_n)\in [0,1]\times [0,1]$, we single out those that are in the rectangle $\mathcal{R}_{q,n}(\mathbf{0})$. We identify them by their indices, which we collect into the set
\[
\mathcal{M}_{q,n}:=\big\{ i: (u_i,v_i)\in \mathcal{R}_{q,n}(\mathbf{0}) \big\}
\subseteq \{1,\dots , n\}.
\]
Let $m_{q,n}:=\#(\mathcal{M}_{q,n})$ denote the cardinality of $\mathcal{M}_{q,n}$, that is, the number of those pairs $(u_i,v_i)$ that are residing in $\mathcal{R}_{q,n}(\mathbf{0})$. Obviously, $m_{q,n}\le n$ and so when we make the assumption that $m_{q,n}$ is sufficiently large, we also implicitly say that the underlying sample size $ n$ must be large. The quantity
\[
\Pi_n^*(q):={1\over n}\sum_{i=1}^n \mathds{1}\big\{(u_i,v_i)\in \mathcal{R}_{q,n}(\mathbf{0}) \big\}={m_{q,n} \over n}
\]
is an empirical proxy for
\[
\Pi^*(q)= \mathbb{P}\big ( (U,V)\in \mathcal{R}_{q}(\mathbf{0})\big ).
\]

\begin{note}\label{note-22}
This paper is long and complex, and thus we have tried -- whenever reasonable -- to avoid pedantic details of the proof. For example, results like $\Pi_n^*(q)$ being a proxy for $\Pi^*(q)$ would normally go without a proof in the current paper, but to illustrate this very instance, suppose that we wish to check the result. Since
\[
\Pi^*_n(q) = \max_{x\in[q^2,1]}C_n(x,q^2/x)\quad \textrm{and} \quad \Pi^*(q) = \max_{x\in[q^2,1]}C(x,q^2/x)
\]
for $q\in(0,1]$, we have
\begin{align*}
|\Pi^*_n(q) - \Pi^*(q)|
&\le \max_{x\in[q^2,1]}|C_n(x,q^2/x) - C(x,q^2/x)|
\\
&\le \sup_{u,v\in[0,1]}|C_n(u,v) - C(u,v)|\stackrel{\mathrm{p}}{\to} 0
\end{align*}
when $n\to\infty$, where $\stackrel{\mathrm{p}}{\to}$ denotes convergence in probability. As we see from the classical results of \citet{KW1958}, and \citet{Kiefer1961}, the convergence holds even almost surely. Much work has been done since those pioneering studies to relax the iid assumption, and we refer to, e.g., \citet{DZ2008}, and \citet{KW2014} for results and references on the topic. The latter studies also provide clues as to why the departure from the iid assumption of \citet{SYZ2020} has been so challenging to develop.
\end{note}

Next, we fix any $m\ge 1$ such that $m\le m_{q,n}$, which holds (recall the message of Note~\ref{note-22}) with as large a probability as desired, assuming that $n$ is sufficiently large.
Then we assign the pairs $(u_i,v_i)$, $i\in \mathcal{M}_{q,n}$, into $\lceil {m_{q,n}}/m\rceil$ disjoint groups so that there can be at most one group with less than $m$ pairs and all the other groups contain exactly $m$ pairs, where $\lceil \cdot \rceil$ is the classical ceiling function. We collect the indices of the grouped pairs into the (disjoint) sets $G_{j,q,n}$, $1\le j\le \lceil {m_{q,n}}/m\rceil$, thus producing a partition of $\mathcal{M}_{q,n}$.

\begin{note}
Choosing an appropriate value of $m$ is a delicate problem in practice. We shall discuss it in great detail when working with simulated and real data later in this paper.
\end{note}

We define the \emph{average block-minima estimator} of the TOMD $\kappa^*$ as follows:
\begin{equation}
\widehat{\kappa}^*_{m_{q,n}}(m,\theta,q)
= \dfrac{1}{\lceil {m_{q,n}}/m\rceil}\sum^{\lceil {m_{q,n}}/m\rceil}_{j=1}\min_{i\in G_{j,q,n}}\dfrac{2T_{\theta}\circ F_{q,\mathcal{M}_{q,n}}(u_i,v_i)}{\log u_i + \log v_i  -2\log q  },
\label{tomd}
\end{equation}
where, for all $u,v\in [0,1]$,
\begin{equation}
F_{q,\mathcal{M}_{q,n}}(u,v)
= {1\over m_{q,n}}\sum_{i\in \mathcal{M}_{q,n}}\mathds{1}\big\{u_i\le u,v_i\le v\},
\label{tomd-cdf}
\end{equation}
and, for all $t\in(0,1]$,
$$
T_{\theta}(t) = \begin{cases}
(t^{\theta} - 1)/\theta & \mbox{when}\quad\theta >0,  \\
\log(t) &  \mbox{when}\quad\theta = 0.
\end{cases}
$$

For comparison, we also consider the TODD $\kappa $. The construction of its estimator follows the ideas of \citet{Gabaix2011}. Namely, let
\[
\mathcal{N}_{q,n} = \big\{ i: (u_i,v_i)\in \mathcal{S}_{q}(\mathbf{0})  \big\} \subseteq \big \{1, \dots, n \big \},
\]
where
\[
\mathcal{S}_q(\mathbf{0})=[0,q]\times [0,q].
\]
Furthermore, let $n_{q,n}:=\#(\mathcal{N}_{q,n})$, which is the cardinality of $\mathcal{N}_{q,n}$. That is, $n_{q,n}$ is the number of pairs $(u_i,v_i)$ in the square $\mathcal{S}_{q }(\mathbf{0})$. For each $i\in \mathcal{N}_{q,n}$, denote
$w_i := q\min\{u^{-1}_i,v^{-1}_i\}$. We order these $w$'s and obtain the order statistics  $w_{1:n_{q,n}}\ge\cdots\ge w_{n_{q,n}:n_{q,n}}$.
An estimator of the TODD $\kappa $ is
\begin{equation}\label{est-ols}
\widehat{\kappa}^{\text{OLS}}_{n_{q,n}} = \dfrac{\sum^{n_{q,n}}_{i=1}\big(\log w_{i:n_{q,n}} - \overline{\log w}~\big)\log(i-0.5)}{\sum^{n_{q,n}}_{i=1}\big(\log w_{i:n_{q,n}} - \overline{\log w}~\big)^2},
\end{equation}
where
$\overline{\log w}$ denotes the average of $\log w_{1},\dots , \log w_{n_{q,n}}$.

\begin{note}
We refer to \citet{Gabaix2011} for an illuminating discussion of why $0.5$ needs to be  subtracted from $i$ under the logarithm sign in the definition of the TODD estimator.
\end{note}

\begin{note}
For a theory of order statistics, whose knowledge is needed to fully appreciated the TODD estimator, as well as other parts of the present paper, we refer to \citet{DN2003}, \citet{ABN2008}, and \citet{ANS2013}.
\end{note}

\section{Justification of the methodology}
\label{application}

In practice we cannot know whether slowly varying functions are present in  equations such as~\eqref{tail eq} and \eqref{max-tail eq}. This poses a challenge. With this in mind, in this section we redevelop the methodology of \citet{SYZ2020} under the uncertainty of the existence of slowly varying functions, but this is just a small part of the innovation that the present paper offers.

An important feature that permeates our following considerations is a conditioning argument, whose main idea is based on the fact that extreme financial losses are those that are below a certain (small) threshold $q>0$, which is often set by convention or regulation. Consequently, we shall deal with conditional copulas below $q$, which are ratios of probabilities. Via equation~\eqref{max-tail eq}, we shall reduce these ratios to quantities like $\ell^*(qu)/\ell^*(q)$, which are close to $1$ when $q>0$ is sufficiently small, due to the assumption that $\ell^*$ is slowly varying at $0$.

When choosing $q>0$ in practice, we should be mindful of the fact that smaller thresholds $q>0$ lead to smaller numbers $m_{q,n}$ of those pairs that are in the rectangle  $\mathcal{R}_{q,n}(\mathbf{0})$. This obviously impedes statistical inference, and so we  need to strike a balance between $q$ and $m_{q,n}$. We shall need to give a considerable thought to this matter when working with real data (see, in particular, Appendix~\ref{append-2}).

\subsection{Estimating the TOMD $\kappa^*$}
\label{sect-42}

Relying on the intuition developed by~\citet{SYZ2020}, we start with equation~\eqref{max-tail eq} and express the conditional maximal tail probability as
\begin{align}
{\Pi^*(qu)\over \Pi^*(q)}
&=
\dfrac{C(\varphi^*(qu),q^2u^2/\varphi^*(qu))}{C(\varphi^*(q),q^2/\varphi^*(q))}
\notag
\\
&= \dfrac{\ell^*(qu)(qu)^{\kappa^*}}{\ell^*(q)q^{\kappa^*}}
\notag
\\
&\approx u^{\kappa^*}
\label{approx-10}
\end{align}
when $ 0< q\approx 0$, due to statement~\eqref{ratio-ells}.
Hence, we have the approximate bound
\begin{equation}
\dfrac{C(\varphi(qu),q^2u^2/\varphi(qu))}{C(\varphi^*(q),q^2/\varphi^*(q))} \lessapprox u^{\kappa^*}
\label{approx-10a}
\end{equation}
that holds for every $\varphi\in \mathcal{A}$. With $\widetilde{u} := \varphi(qu)/\varphi^*(q)$ and $\widetilde{v} := u^2\varphi^*(q)/\varphi(qu)$,
we turn bound~\eqref{approx-10a} into
\[
 \dfrac{C(\varphi^*(q)\widetilde{u},q^2\widetilde{v}/\varphi^*(q))}{C(\varphi^*(q),q^2/\varphi^*(q))} \lessapprox  u^{\kappa^*} .
\]
Since $u^{\kappa^*} = (\widetilde{u}\widetilde{v})^{\kappa^*/2}$, this leads to
\begin{equation}\label{bound-9}
\kappa^* \lessapprox  \dfrac{2\log F^*_{q}(\widetilde{u},\widetilde{v})}{\log \widetilde{u} + \log \widetilde{v}},
\end{equation}
where $F_q^*:[0,1]\times [0,1] \to [0,1]$ is defined by
\begin{equation}\label{def-fq}
F^*_{q}(u,v) = \dfrac{C(u\varphi^*(q),vq^2/\varphi^*(q))}{C(\varphi^*(q),q^2/\varphi^*(q))}.
\end{equation}
This gives us a theoretical foundation for building an empirical estimator for the TOMD $\kappa^*$.

Namely, given pseudo observations $(u_1,v_1), \dots , (u_n,v_n)\in [0,1]\times [0,1]$ and also a small fixed $q>0$, the TOMD $\kappa^*$ is estimated using the following four-step procedure:
\begin{enumerate}[(1)]
\item
Using all the pairs available in the unit square $[0,1]\times [0,1]$,
we estimate $\varphi^*(q)$ by maximizing $C_n(x,q^2/x)$ with respect to $x\in[q^2,1]$.
\item
We extract the set $\mathcal{P}_{q,n}$ of those pairs $(u_k,v_k)$ that are in the rectangle $\mathcal{R}_{q,n}(\mathbf{0})$, and then collect the indices of the pairs into the set $\mathcal{M}_{q,n}$, whose cardinality we denote by $m_{q,n}$.
\item
We randomly assign the pairs of $\mathcal{P}_{q,n}$ into $\lceil m_{q,n}/m\rceil$ disjoint groups of pairs, whose indices partition the set $\mathcal{M}_{q,n}$ into groups $G_{1,q,n},\ldots, G_{\lceil m_{q,n}/m\rceil, q,n}$ with at most one of them having fewer than $m$ elements, while all the other groups having exactly $m$ elements.
\item
Finally, we compute the average block-minima estimator of the TOMD $\kappa^*$ using formula~\eqref{tomd}.
\end{enumerate}

To appreciate formula~\eqref{tomd} in view of the above procedure, note that, for $i\in\mathcal{M}_{q,n}$,
\begin{align*}
F_{q,\mathcal{M}_{q,n}}(u_i,v_i)
&=\dfrac{1}{m_{q,n}}\#\big \{k\in \mathcal{M}_{q,n}: u_k\le u_i, v_k\le v_i\big \}
\\
&=F^*_{q,\mathcal{M}_{q,n}}(\widetilde{u}_i,\widetilde{v}_i),
\end{align*}
where
\begin{equation}\label{def-fqm}
F^*_{q,\mathcal{M}_{q,n}}(u,v)
= \dfrac{1}{m_{q,n}}\sum_{k\in\mathcal{M}_{q,n}}
\mathds{1}\big \{\widetilde{u}_k\le u, \widetilde{v}_k\le v \big \}
\end{equation}
with
\[
\widetilde{u}_i = {u_i \over \varphi^*_n(q)}
\quad \textrm{and} \quad
\widetilde{v}_i ={v_i\varphi^*_n(q) \over q^2}.
\]
Hence, estimator~\eqref{tomd} can be rewritten as
\begin{equation}\label{Cqstar}
\widehat{\kappa}^*_{m_{q,n}}(m,\theta,q) = \dfrac{1}{\lceil m_{q,n}/m\rceil}\sum^{\lceil m_{q,n}/m\rceil}_{j=1}\min_{i\in G_{j,q,n}}\dfrac{2T_{\theta}\circ F^*_{q,\mathcal{M}_{q,n}}(\widetilde{u}_i,\widetilde{v}_i)}{\log \widetilde{u}_i + \log \widetilde{v}_i}.
\end{equation}

\subsection{Estimating the TODD $\kappa$}
\label{sect-41}

To provide an effective comparison of the already discussed estimation of the TOMD $\kappa^*$ with the estimation of the TODD $\kappa$, we next follow -- modulus some modifications -- the spirit of Section~\ref{sect-42}. Namely, in view of equation~\eqref{tail eq}, we have
\begin{equation}
\dfrac{C(qu,qu)}{C(q,q)} = \dfrac{\ell(qu)(qu)^{\kappa}}{\ell(q)q^{\kappa}}\approx u^{\kappa}
\label{approx-11}
\end{equation}
when $ 0< q\approx 0$, due to statement~\eqref{ratio-ells}. To see an analogy between equations~\eqref{approx-11} and \eqref{approx-10}, we set $\varphi^*(u)=u$ in statement~\eqref{approx-10}. Next, we rewrite approximate equation~\eqref{approx-11} as
\begin{equation}\label{abc-0}
\mathbb{P}\big (\max\{U,V\}\le q u \mid \max\{U,V\}\le q \big ) \approx u^{\kappa}.
\end{equation}
With the notation $z:=u^{-1}$, equation~\eqref{abc-0} turns into
\begin{equation}\label{TO: tail copula, w prob}
\mathbb{P}\big (W \ge z \mid W \ge 1 \big ) \approx z^{-\kappa},
\end{equation}
where $W=q\min\{U^{-1},V^{-1}\}$. To implement approximate equation~\eqref{TO: tail copula, w prob} on data, we produce its empirical version, and for this, we follow the ideas of \citet{Gabaix2011}. Namely, in terms of the ordered $w$'s, which are $w_{1:n_{q,n}}\ge\cdots\ge w_{n_{q,n}:n_{q,n}}$, the empirical version of statement~\eqref{TO: tail copula, w prob} becomes
\begin{equation}\label{abc-1}
i/n_{q,n} \approx w^{-\kappa}_{i:n_{q,n}}
\end{equation}
for $i=1,\ldots,n_{q,n}$.
Taking the logarithms of the two sides of equation~\eqref{abc-1}, we obtain
\[
\log i \approx \log n_{q,n} - \kappa\log w_{i:n_{q,n}}.
\]
From this, we obtain an estimator of $\kappa$ as the slope of the least squares regression line fitted to the scatterplot of the pairs
$\big (\log(i-0.5), \log w_{i:n_{q,n}} \big )$, $i=1,\dots,n_{q,n}$. This gives rise to estimator~\eqref{est-ols}.

\subsection{A theoretical justification of the TOMD estimator}
\label{append-0}

We can and thus do use all the available pseudo observations $(u_1,v_1), \dots , (u_n,v_n)\in [0,1]\times [0,1]$ to get estimates of $C$, $\varphi^*(q)$, $\mathcal{R}_q(\mathbf{0})$, and $\Pi^*(q)$. Since in our applications the sample size $n$ is large, it is reasonable to assume that the estimates are close -- or as close as we can possibly make them -- to their population versions, and we can thus adopt the same notations for all of them. That is, in what follows we simply skip the subscript $n$ from the estimators of $C$, $\varphi^*(q)$, $\mathcal{R}_q(\mathbf{0})$, and $\Pi^*(q)$.

The real challenge is that the number $m_{q}$ of the observed pairs in the estimated rectangle $\mathcal{R}_q(\mathbf{0})$ is relatively small, and thus their variability matters. We thus face the question of whether $m_{q}$ is sufficiently large in order to make a reliable statistical decision. To answer this question, we next theoretically explore the convergence of the simplified TOMD estimator
\begin{equation}
\widehat{\kappa}^*_{m_{q}}(m,0,q)
= \dfrac{1}{\lceil {m_{q}}/m\rceil}\sum^{\lceil {m_{q}}/m\rceil}_{j=1}\min_{i\in G_{j,q}}\dfrac{2T_{0}\big( C(U_{i,q},V_{i,q})/\Pi^*(q)\big)}{\log U_{i,q} + \log V_{i,q}  -2\log q  }
\label{tomd-2pp}
\end{equation}
when $m_{q}\to \infty $, where $(U_{i,q},V_{i,q})$, $i\in \mathcal{M}_{q}$, are iid random pairs, each following the conditional cdf $F_q:\mathcal{R}_q(\mathbf{0}) \to [0,1]$ defined by
\begin{align*}
F_q(u,v)
:=&\mathbb{P}\big (U\le u,V \le v \mid (U,V) \in \mathcal{R}_q(\mathbf{0}) \big )
\\
=& {1 \over \Pi^*(q)} C\big (\min\{u, \varphi^*(q)\},\min\{v, q^2/\varphi^*(q)\}\big ).
\end{align*}
Thus, $F_q(u,v)=C(u,v) /\Pi^*(q)$  for all $(u,v)\in \mathcal{R}_q(\mathbf{0})$.
By the law of large numbers, when $m_{q}\to\infty$, we have
\begin{align}
\widehat{\kappa}^*_{m_{q}}(m,0,q)
\stackrel{\mathrm{p}}{\to}
\kappa^*(m,0,q)
:=\mathbb{E}\left[\min_{i\in G_{1,q}}\dfrac{2T_{0}\big( C (U_{i,q},V_{i,q})/\Pi^*(q)\big)}{\log U_{i,q} + \log V_{i,q}-2\log q }\right].
\label{tomd-2app}
\end{align}

At first sight, assuming independence of the pairs $(U_{i,q},V_{i,q})$, $i\in \mathcal{M}_{q}$, might be alarming, given that we aim at analyzing (dependent) time series data. However, since for reasons of mitigating the influence of slowly varying function, we use only those observed pairs that are in a (small) neighbourhood of $\mathbf{0}$. This gives rise to nearly independent data, as we shall see in Section~\ref{append-2}. There is, of course, a gap between independent variables and nearly independent ones, such as white noise, but our estimator, with its block structure, is fairly robust with respect to dependence, as we shall see in  Section~\ref{appl-simulations}. Hence, we can comfortably assume that the random pairs $(U_{i,q},V_{i,q})$, $i\in \mathcal{M}_{q}$, are iid, for the sake of convincing ourselves that, in principle, the TOMD estimator does work. (As we have already noted, a mathematically pedantic theory with all the technical details is simply too lengthy for a single paper.)

The following theorem shows that by choosing $m\ge 1$ and $q>0$ appropriately, we can make $\kappa^*(m,0,q)$ as close to the TOMD $\kappa^*$ as desired, and thus, by statement~\eqref{tomd-2app}, the estimator $\widehat{\kappa}^*_{m_{q}}(m,0,q)$ can be made as close to the TOMD $\kappa^*$ as desired.

\begin{theorem}\label{thm: essinf}
Let $q\in (0,1]$ be any, and let the cdf $F_q^*$ satisfy the bound
\begin{equation}\label{pqd}
F_q^*(u,v)\ge uv\quad \textrm{for all}\quad u,v\in [0,1].
\end{equation}
Furthermore, let the functions $\varphi^*(u)$ and $\psi^*(u) := u^2/\varphi^*(u)$ be   strictly increasing. Then $\kappa^*(m,0,q)$ can be made as close to $ \kappa^*$ as desired by taking sufficiently large $m\ge 1$ and sufficiently small $q>0$.
\end{theorem}

Theorem~\ref{thm: essinf} is pivotal for developing practical statistical inference for the TOMD estimator as it justifies the use of $\widehat{\kappa}^*_{m_{q}}(m,0,q)$ for estimating $ \kappa^*$ at a prescribed (fixed) margin of error and increasing (with respect to the sample size) confidence level. For details on this path of reasoning and its practicality, although in a different context, we refer to \citet{GSZ2022a}, with further details and contrasts between this and the classical paths provided by  \citet{GSZ2022b}. Hence, in the present paper we do not attempt to develop confidence intervals with fixed confidence levels and shrinking margins of error, as they would be based on the asymptotic distribution of the TOMD estimator. Although knowing the asymptotic distribution is beneficial, deriving such results requires considerable space. The same also applies to the other estimators that we discuss in the current paper.

The proof of Theorem~\ref{thm: essinf} is rather technical, and we have relegated it to Appendix~\ref{proof-of-th31}. As to the practical side of the theorem, the biggest challenge when using it is to convince oneself in the plausibility of condition~\eqref{pqd}. We devote Appendices~\ref{append-3a} and~\ref{append-3} to demonstrate how this can be done in the case of real data by employing  statistical tests. In this regard, we note that condition~\eqref{pqd} can be interpreted as the requirement that the cdf $F_q^*$ does not dip below the independence copula
\begin{equation}\label{indepent-copula}
C^{\perp}(u,v):=uv
\end{equation}
at any point $u,v\in [0,1]$. Specifically, in Appendix~\ref{append-3a} we test the hypothesis $F_q^*\ge C^{\perp}$ (i.e., condition~\eqref{pqd}) while in Appendix~\ref{append-3} we tackle the ``boundary'' case $F_q^*=C^{\perp}$.

\subsection{Insights into $F_q^*$: two examples}
\label{illustrate-fq}

In this section we develop insights into the form of $F_q^*$ in the case of the Marshall-Olkin and the generalized Clayton copulas. In addition to $F_q^*$, we also discuss the validity of bound~\eqref{pqd} and also calculate the limit
\[
F_0^*(u,v) := \lim_{q\downarrow 0}F_q^*(u,v), \quad u,v\in[0,1].
\]

\begin{example}
The Marshall-Olkin (M-O) copula $C_{a,b}:[0,1]\times [0,1] \to [0,1]$ is defined by \citep{mo1967}
\[
C_{a,b}(u,v) = \min(u^{1-a}v, uv^{1-b})
\]
with parameters $a,b\in[0,1]$.  For every $q\in[0,1]$, we have \citep{Furman2015}
$\varphi^*(q) = q^{2b/(a+b)} $ and $ \psi^*(q) = q^{2a/(a+b)}$,
and so
$\Pi^*(u) = u^{\kappa^*}$ with the TOMD $ \kappa^* = 2 - 2ab/(a+b) $.
Consequently, definition~\eqref{def-fq} immediately gives
\begin{align*}
F_q^*(u,v)
&= \dfrac{\min\{(q^{2b/(a+b)}u)^{1-a}(q^{2a/(a+b)}v), (q^{2b/(a+b)}u)(q^{2a/(a+b)}v)^{1-b}\}}{q^{2-2ab/(a+b)}}
\\
&= \min\{u^{1-a}v,uv^{1-b}\} ,
\end{align*}
which is equal to $C_{a,b}(u,v)$. Hence, $F_0^*=C_{a,b}$. This implies that both $F_q^*$ and $F_0^*$ satisfy bound~\eqref{pqd}, because the M-O copula $C_{a,b}$ is PQD. For additional information on the M-O copula and its applications to financial risk management, we refer to, e.g., \citet{afv2010,afv2016}.
\end{example}

\begin{example}
The generalized Clayton (GC) copula $C_{\gamma_0,\gamma_1}:[0,1]\times [0,1] \to [0,1]$ is defined by
\begin{equation}\label{def-gcc}
C_{\gamma_0,\gamma_1}(u,v) = u^{\gamma_1/\gamma^*}\big(u^{-1/\gamma^*} + v^{-1/\gamma_0} - 1\big)^{-\gamma_0}
\end{equation}
with $\gamma_0>0$, $\gamma_1\ge0$, and $\gamma^* := \gamma_0 + \gamma_1$. For details on this copula and its applications to financial risk management, we refer to \citet{sf2017,sf2018}.
In view of formula~\eqref{def-gcc}, we have
\begin{align*}
F^*_q(u,v)
&= u^{\gamma_1/\gamma^*} \bigg (\dfrac{(\varphi^*(q)u)^{-1/\gamma^*} + (\psi^*(q)v)^{-1/\gamma_0} - 1}{\varphi^*(q)^{-1/\gamma^*} + \psi^*(q)^{-1/\gamma_0} - 1}\bigg )^{-\gamma_0} \\
&= u \bigg (\dfrac{\varphi^*(q)^{-1/\gamma^*} + (\psi^*(q)v)^{-1/\gamma_0}u^{1/\gamma^*} - u^{1/\gamma^*}}{\varphi^*(q)^{-1/\gamma^*} + \psi^*(q)^{-1/\gamma_0} - 1}\bigg )^{-\gamma_0} \\
&\ge  u \bigg (\dfrac{\varphi^*(q)^{-1/\gamma^*} + (\psi^*(q)v)^{-1/\gamma_0} - 1}{\varphi^*(q)^{-1/\gamma^*} + \psi^*(q)^{-1/\gamma_0} - 1}\bigg )^{-\gamma_0} \\
&\ge  u \bigg (\dfrac{(\psi^*(q)v)^{-1/\gamma_0}}{\psi^*(q)^{-1/\gamma_0}}\bigg )^{-\gamma_0} = uv.
\end{align*}
This shows that $F^*_q$ satisfies bound~\eqref{pqd}.

Note that  $\varphi^*(q)$ satisfies the equation \citep[equation~(6.1)]{Furman2015}
$$
\varphi^*(q)^{-1/\gamma_0}(\varphi^*(q)^{-1/\gamma^*} - (\gamma_1/\gamma^*)) = (1 - (\gamma_1/\gamma^*))q^{-2/\gamma_0},
$$
which reduces to
$
\varphi^*(q)^{-1/\gamma^*} - (\gamma_1/\gamma^*) = (1 - (\gamma_1/\gamma^*))\psi^*(q)^{-1/\gamma_0}
$
and gives
\begin{align*}
F_q^*(u,v) &= u^{\gamma_1/\gamma^*}\bigg (\dfrac{((1 - (\gamma_1/\gamma^*))\psi^*(q)^{-1/\gamma_0}+(\gamma_1/\gamma^*))u^{-1/\gamma^*} + (\psi^*(q)v)^{-1/\gamma_0} - 1}{(2 - (\gamma_1/\gamma^*))\psi^*(q)^{-1/\gamma_0} - (\gamma_0/\gamma^*)}\bigg )^{-\gamma_0}
\\
&\to u^{\gamma_1/\gamma^*}\bigg (\dfrac{(1 - (\gamma_1/\gamma^*))u^{-1/\gamma^*} + v^{-1/\gamma_0}}{2 - (\gamma_1/\gamma^*)}\bigg )^{-\gamma_0}
\\
&=: F_0^*(u,v)
\end{align*}
when $q\downarrow 0$.

We next show that $F_0^*$ is related the TOMD $\kappa^*$ via the bound
\begin{equation}\label{relat-gcc}
F_0^*(u,v)\le (uv)^{\kappa^*/2} ,
\end{equation}
which holds for all $u,v\in [0,1]$, where \citep{Furman2015}
\begin{equation}\label{tomd-gcc}
\kappa^* = 1 + \dfrac{\gamma_1}{\gamma_1 + 2\gamma_0}.
\end{equation}
To begin, we note that, for any $q\in(0,1)$, maximizing $F_0^*(x,q^2/x) $ over $x\in[q^2,1]$ is equivalent to maximizing
$$x\mapsto \dfrac{\gamma_1}{\gamma^*}\log x - \gamma_0\log\bigg(\dfrac{(1 - (\gamma_1/\gamma^*))x^{-1/\gamma^*}+ (q^2/x)^{-1/\gamma_0}}{2 - (\gamma_1/\gamma^*)}\bigg).$$
The first-order condition is
$$\dfrac{\gamma_1}{\gamma^*x} - \dfrac{\gamma_0(1 - (\gamma_1/\gamma^*))(-1/\gamma^*)x^{-1/\gamma^*-1}+ q^{-2/\gamma_0}x^{1/\gamma_0-1}}{(1 - (\gamma_1/\gamma^*))x^{-1/\gamma^*}+ (q^2/x)^{-1/\gamma_0}} = 0,$$
which reduces to
$x^{-1/\gamma^*} = (q^2/x)^{-1/\gamma_0}$
and gives the solution
$x^* =q^{2\gamma^*/(\gamma^*+\gamma_0)}.$
Consequently,
\begin{align*}
\max_{x\in[q^2,1]}F_0^*(x,q^2/x)
&= F_0^*(q^{2\gamma^*/(\gamma^*+\gamma_0)},q^{2\gamma_0/(\gamma^*+\gamma_0)})
\\
&= q^{2\gamma_1/(\gamma^*+\gamma_0)}  q^{2\gamma_0/(\gamma^*+\gamma_0)}
\\
&= q^{1 + \gamma_1/(\gamma_1+2\gamma_0)}.
\end{align*}
By \citet[equation~(6.2)]{Furman2015}, this implies bound~\eqref{relat-gcc}.
\end{example}

\section{An illustrative simulation study}
\label{appl-simulations}

The definition of the estimator $\widehat{\kappa}^*_{m_{q}}(m,\theta,q)$ does not require observations $(u_i,v_i)$, $i\in \mathcal{M}_q$, to arise from independent pairs $(U_i,V_i)$, although \citet{SYZ2020} established its consistency and other statistical properties under this  assumption. Given that we are to apply the estimator on pairs arising from time series, we wish to check the estimator's robustness with respect to dependent data. The time series models that are suitable for financial instruments -- such as foreign currency exchange rates, stock market indices, and treasury notes -- are complex. They follow, e.g., the ARIMA model for the conditional mean and the GARCH, or some other heteroscedastic, model for the conditional variance.

Since we are concerned with the co-movements of extreme losses, which are few and far between inside the original time series, the extreme losses follow models  fairly close to a white noise. We shall check this conjecture in Appendix~\ref{append-2} using a number of portmanteau tests. Hence, in order to check the performance of the TOMD estimator when the iid assumption is slightly violated, we shall conduct a simulated experiment when the observed pairs $(u_i,v_i)$ arise from a time series which is not too far away from being a bivariate  white noise  \citep{bjrl2015}. Specifically, we shall next describe a procedure for simulating random pairs whose intra-pair dependence is governed by the generalized Clayton copula and the inter-pair dependence arises from the AR(1) time-series model.

To generate $(U_i,V_i)\sim C_{\gamma_0,\gamma_1}$, we start with the conditional cdf of $U$ given $V=v$, which has the expression
\begin{align*}
\mathbb{P}(U\le u\mid V = v)
&= \dfrac{\partial}{\partial v}C_{\gamma_0,\gamma_1}(u,v)
\\
&= u^{\gamma_1/\gamma^*}\big(u^{-1/\gamma^*}v^{1/\gamma_0} + 1 - v^{1/\gamma_0}\big)^{-\gamma_0-1}.
\end{align*}
With the notation $z = u^{-1/\gamma^*} - 1$, this yields
\begin{align*}
\mathbb{P}\big (U\le (z+1)^{-\gamma^*}\mid V = v\big )
&= (z+1)^{-\gamma_1}\big(zv^{1/\gamma_0} + 1\big)^{-\gamma_0-1}
\\
&= \mathbb{P}(Y>z)\mathbb{P}(X>z\mid V = v)
\\
&=\mathbb{P}\big(\min\{X,Y\}>z\mid V = v\big ),
\end{align*}
where the random variables $Y\sim\text{Lomax}(\gamma_1,1)$ and $[X\mid V=v] \sim \text{Lomax}(\gamma_0 + 1, v^{-1/\gamma_0})$ are independent.
Hence, to simulate a stationary sequence $(U_i,V_i)\sim C_{\gamma_0,\gamma_1}$, $i\in \mathbb{Z}$,  we:
\begin{enumerate}[(1)]
\item\label{step1} generate $Z_i\sim F$ using a time series model;
\item\label{step2} set $V_i := F(Z_i)$;
\item\label{step3} generate independent $X_i\sim\text{Lomax}(\gamma_0 + 1, V^{-1/\gamma_0}_i)$ and $Y_i\sim\text{Lomax}(\gamma_1,1)$;
\item\label{step4} calculate $U_i = (1 + \min\{X_i,Y_i\})^{-\gamma^*}$.
\end{enumerate}

For step~\eqref{step1}, we simulate $Z_i$'s using the strictly stationary and causal AR(1) time series
\begin{equation}\label{time-ex}
Z_i = \phi Z_{i-1} + \varepsilon_i,\quad (\varepsilon_i)\stackrel{\mathrm{iid}}{\sim}\mathcal{N}(0,1),
\end{equation}
where $\phi \in (-1,1)$ regulates the departure of the sequence $(Z_i)$ from the white noise $(\varepsilon_i)$; if $\phi=0$, then they coincide. Since we want  $Z_i$'s to carry some dependence, we set $\phi=0.6$, and so the standard deviation in this case becomes $\sigma = 1/\sqrt{1 - \phi^2}= 1.25$ and the autocovariance $\phi^{n}/(1-\phi^2)$ for all $n\ge 1$.

Following steps~\eqref{step1}--\eqref{step4}, we simulate $\{(u_i,v_i):1\le i \le n:= 500,000\}$ one thousand times, where each pair $(u_i,v_i)$ arises from the generalized Clayton copula $C_{\gamma_0,\gamma_1}$ with the parameter choices specified in Table~\ref{Simu: 5per}. %
\begin{table}[h!]
\centering
\begin{tabular}{cccccc}
\hline\hline
$(\gamma_0,\gamma_1)$ & $\kappa^*$  & Mean & StDev & A-D & C-vM  \\
\hline
$(0.1, 0.8)$ & $1.8$ & $1.7920$ &  $0.0381$ & $0.7136$ & $0.6853$ \\
$(0.4,0.8)$  & $1.5$ & $1.5084$ &  $0.0221$ & $0.6982$ & $0.6217$ \\
$(0.4,0.2)$  & $1.2$ & $1.2090$ &  $0.0127$ & $0.9758$ & $0.9352$ \\
\hline
\end{tabular}
\caption{Summary of simulation results when $q=0.05$.}
\label{Simu: 5per}
\end{table}
The TOMD $\kappa^*$ is calculated using formula~\eqref{tomd-gcc}.

\begin{figure}[h!]
    \centering
    \begin{subfigure}[b]{0.3\textwidth}
        \includegraphics[width=\textwidth]{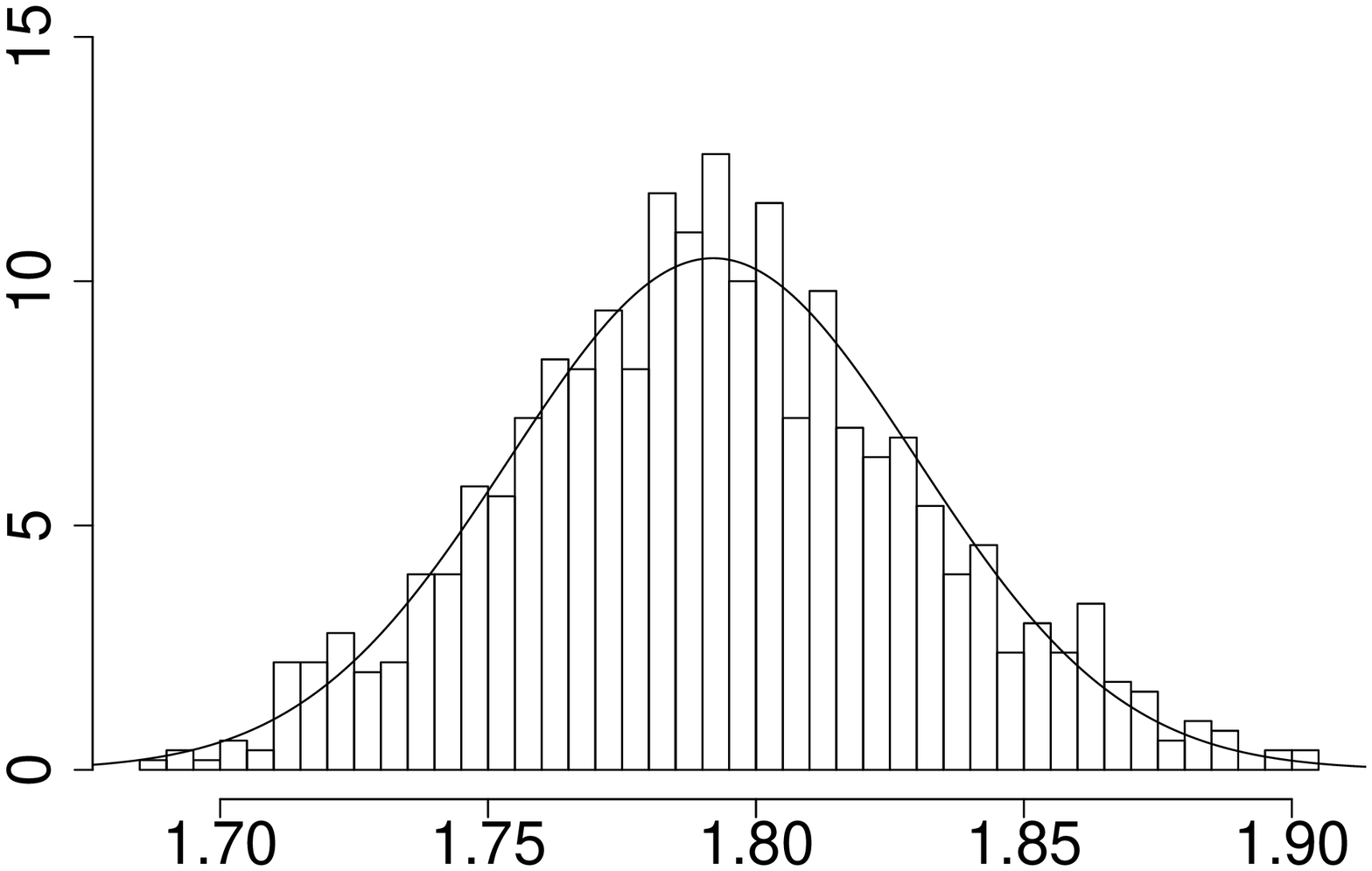}
        \caption{$\kappa^*=1.8$}
    \end{subfigure}
\quad
    \begin{subfigure}[b]{0.3\textwidth}
        \includegraphics[width=\textwidth]{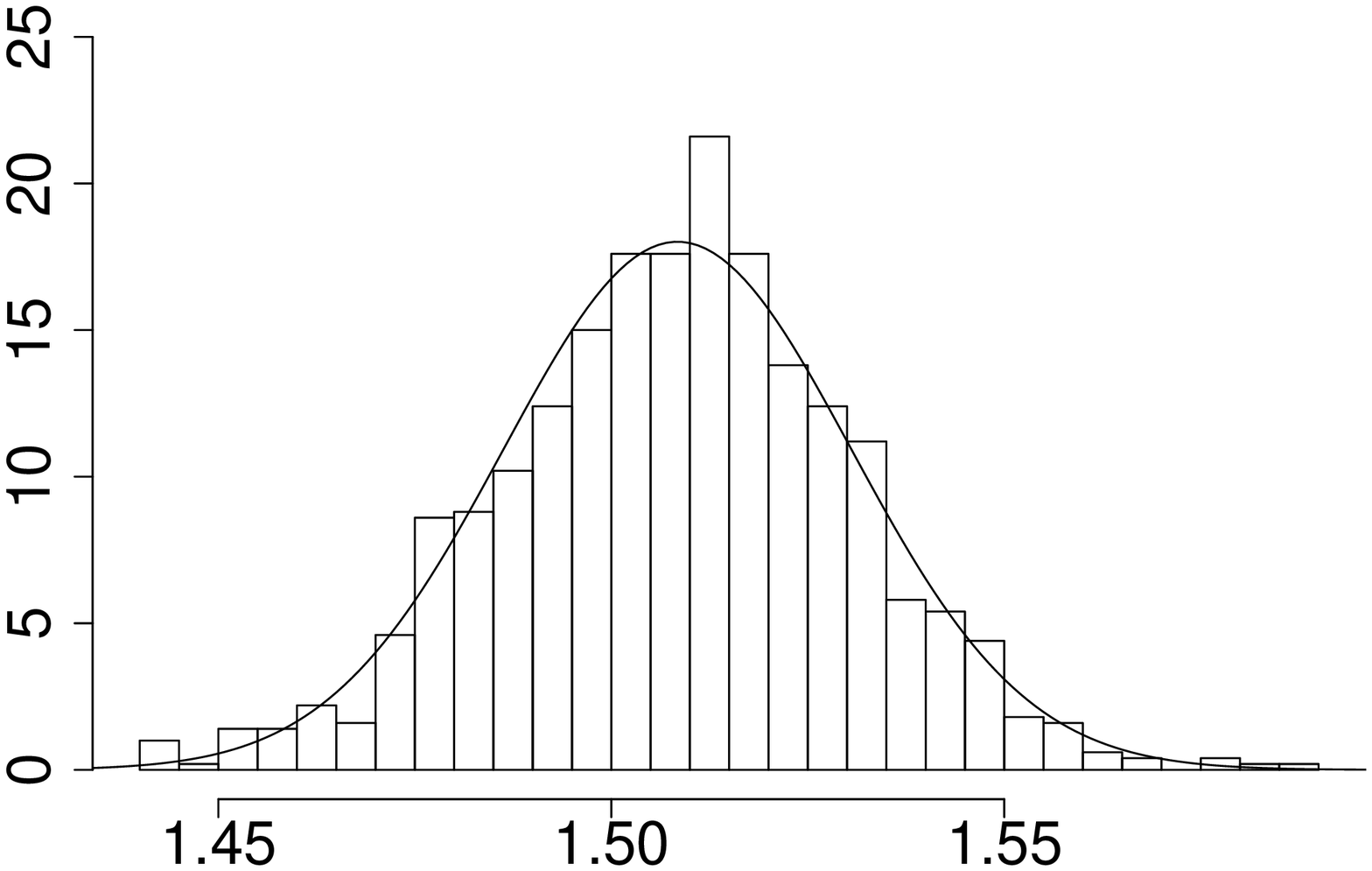}
        \caption{$\kappa^*=1.5$}
    \end{subfigure}
\quad
    \begin{subfigure}[b]{0.3\textwidth}
        \includegraphics[width=\textwidth]{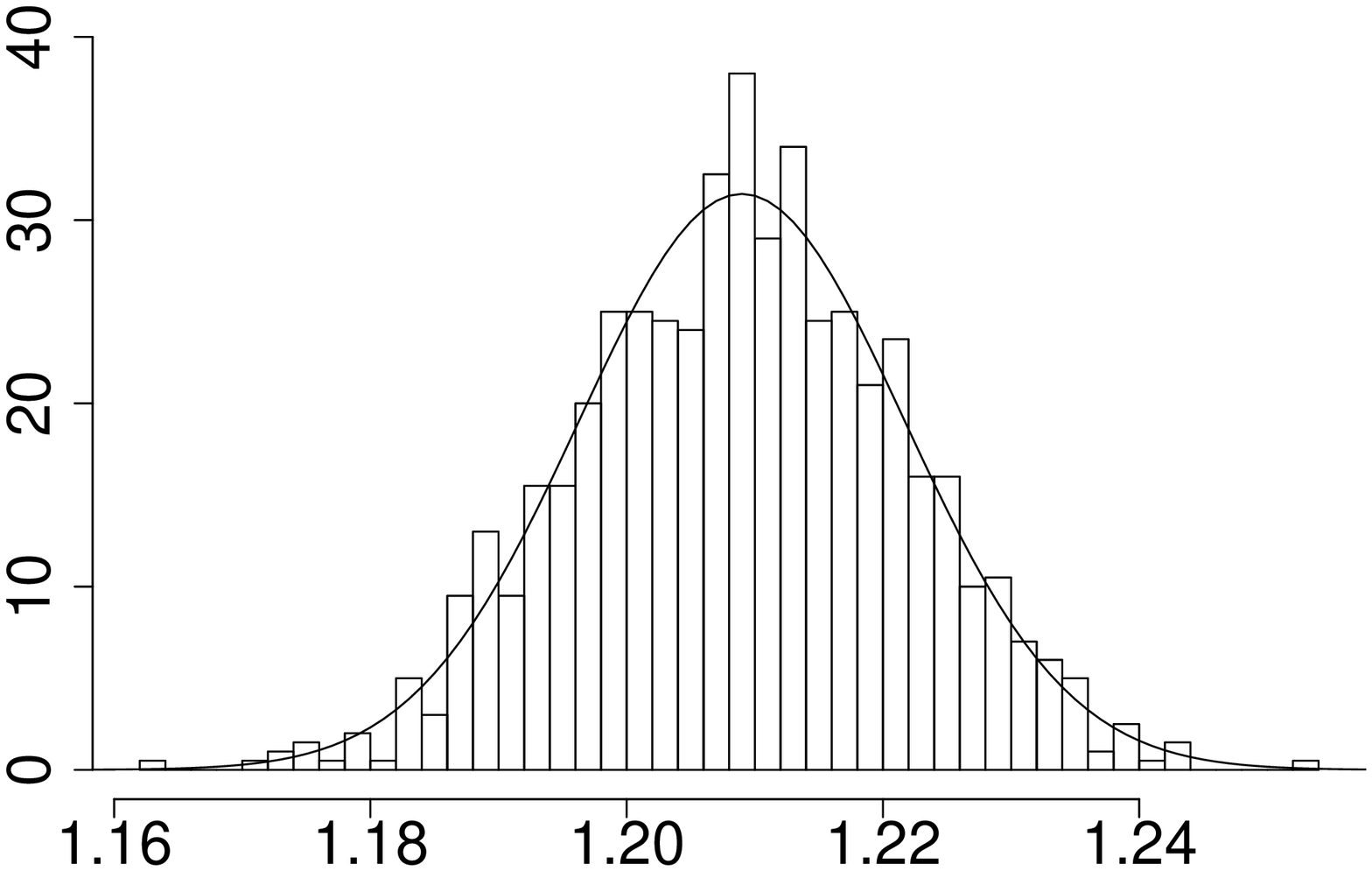}
        \caption{$\kappa^*=1.2$}
    \end{subfigure}
    \caption{Fits of simulated $\widehat{\kappa}^*_{m_{q}}(5,\theta_0,0.05)$ to the normal distribution when $q=0.05$.}
    \label{Fig: 5per}
\end{figure}

Table~\ref{Simu: 5per} also contains several summary statistics when $m = 5$, which is the number of groups $G_{j,q}$, and $q = 0.05$, which is the threshold that serves our working definition of ``extreme.'' Note that the reported $p$-values of the Anderson-Darling (A-D) and Cram\'{e}r-von Mises (C-vM) tests for normality retain the null. The fits of these values  to the normal distribution are depicted in Figure~\ref{Fig: 5per}.  %
When simulating, we always set $\theta$ to $\theta_0:=10^{-6}$.

\begin{figure}[h!]
    \centering
    \begin{subfigure}[b]{0.3\textwidth}
        \includegraphics[width=\textwidth]{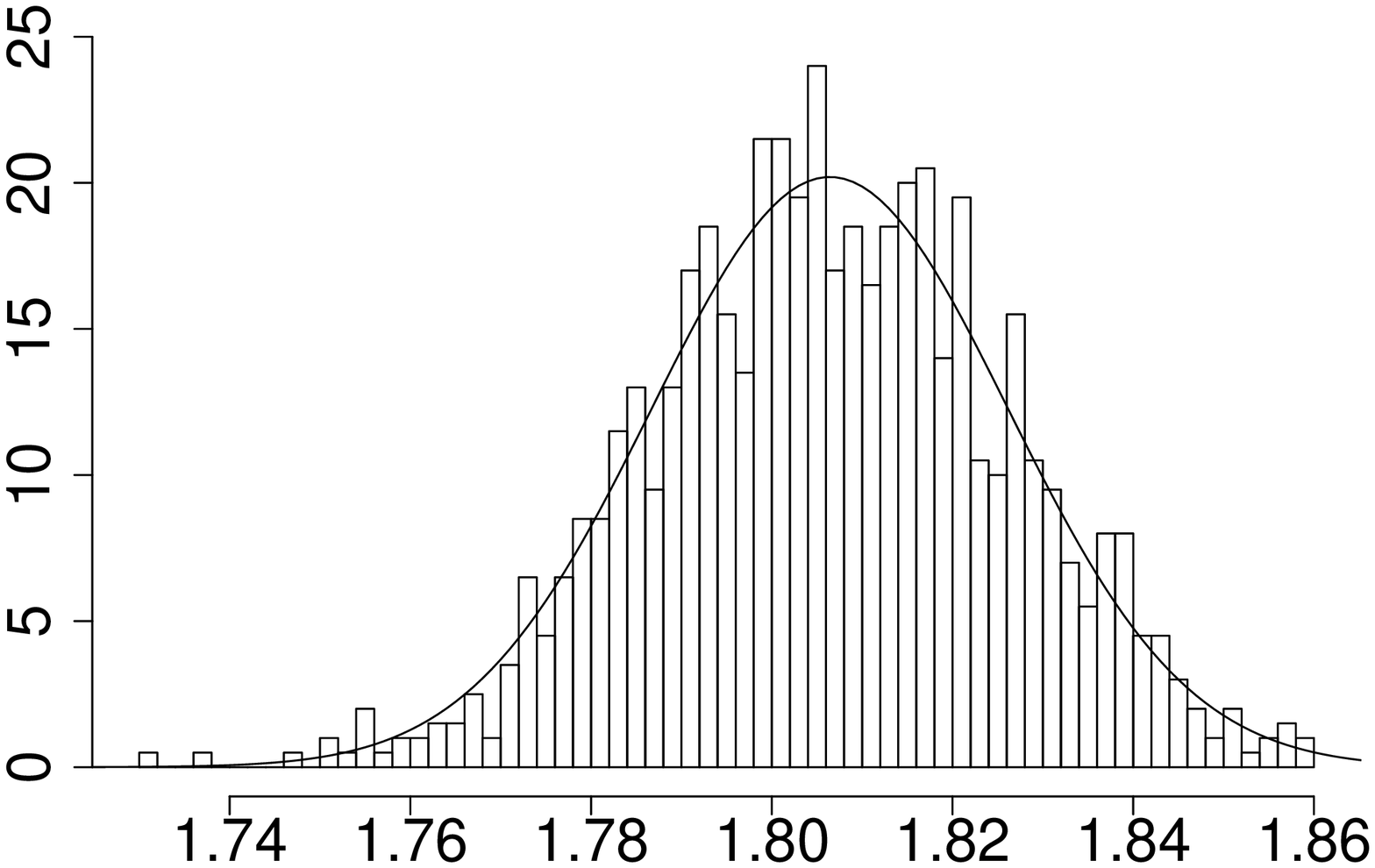}
        \caption{$\kappa^*=1.8$}
        \label{fig:Large5per}
    \end{subfigure}
\quad
    \begin{subfigure}[b]{0.3\textwidth}
        \includegraphics[width=\textwidth]{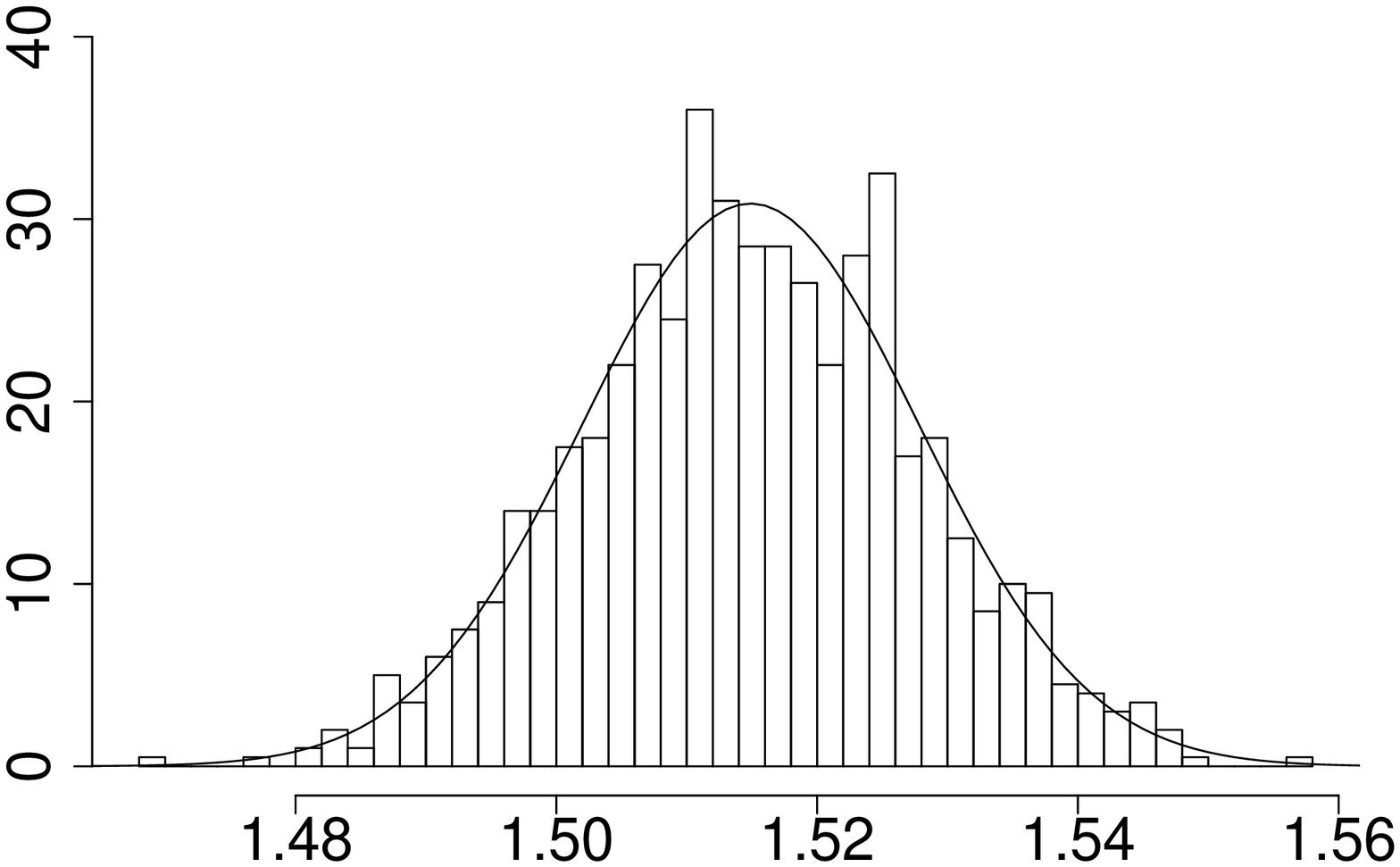}
        \caption{$\kappa^*=1.5$}
        \label{fig:Medium5per}
    \end{subfigure}
\quad
    \begin{subfigure}[b]{0.3\textwidth}
        \includegraphics[width=\textwidth]{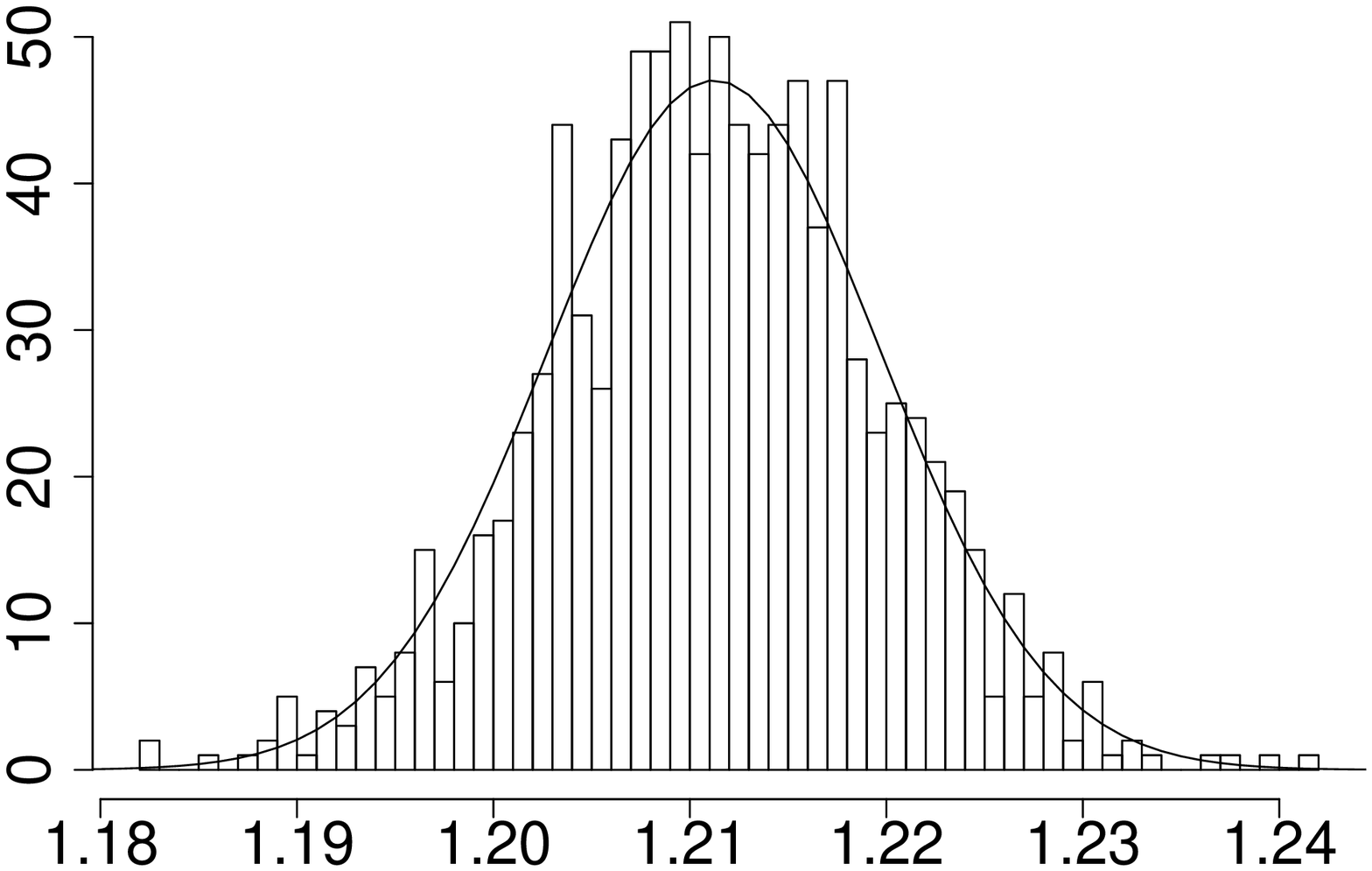}
        \caption{$\kappa^*=1.2$}
        \label{fig:Small5per}
    \end{subfigure}
    \caption{Fits of simulated $\widehat{\kappa}^*_{m_{q}}(5,\theta_0,0.1)$ to the normal distribution when $q=0.1$.}
    \label{Fig: 10per}
\end{figure}

For comparison, we use the same parameters $m$ and $(\gamma_0,\gamma_1)$  but now enlarge the threshold to $q = 0.1$. This increases the number of pairs in the rectangle $\mathcal{R}_q(\mathbf{0})$. Summary statistics are reported in Table~\ref{Simu: 10per}, %
\begin{table}[h!]
\centering
\begin{tabular}{cccccc}
\hline\hline
$(\gamma_0,\gamma_1)$ & $\kappa^*$  & Mean & StDev & A-D & C-vM  \\
\hline
$(0.1, 0.8)$ & $1.8$ & $1.8064$ &  $0.0197$ & $0.9900$ & $0.9948$ \\
$(0.4,0.8)$  & $1.5$ & $1.5149$ &  $0.0129$ & $0.9807$ & $0.9514$ \\
$(0.4,0.2)$  & $1.2$ & $1.2112$ &  $0.0085$ & $0.9322$ & $0.9373$ \\
\hline
\end{tabular}
\caption{Summary of simulation results when $q=0.1$.}
\label{Simu: 10per}
\end{table}
with the A-D and C-vM $p$-values retaining the null of normality. The fits of the simulated values to the normal distribution are depicted in Figure~\ref{Fig: 10per}.  %

\section{Extreme co-movements of financial instruments}
\label{appl-0}

We now explore extreme co-movements of
\begin{itemize}
\item
exchange rates of currencies (CAD/USD, GBP/USD, and JPY/USD)
\item
stock market indices (Dow Jones, S\&P 500, and NASDAQ)
\item
diverse financial instruments (JPY/USD, 10-year Treasury, and NASDAQ)
\end{itemize}
during various periods of time, which have been determined by data availability and/or the date at which we conducted their analyses. Since several data points were missing, we removed the corresponding values from the other time series relevant to our statistical analysis. For the resulting time series $(x^0_t)$, we then calculated $(x_t)$ defined by
\[
x_t=\log(x^0_t)-\log(x^0_{t-1}),
\]
which are depicted in the left-hand panels of Figures~\ref{fig:Log-returns}--\ref{MIX-fig:Log-returns} in Appendix~\ref{append-2}. We have estimated the TOMD $\kappa^*$ using the procedure described in Section~\ref{sect-42}. To compare, we have also estimated the TODD $\kappa $ using the procedure described in Section~\ref{sect-41}. To measure the difference between the two tail orders, we have calculated the relative difference in percentages:
\[
\text{RD}:=\bigg ({\text{TOMD}\over \text{TODD}}-1\bigg )100\% .
\]
From the theoretical point of view, RD can be either negative or zero, but its empirical version can nevertheless sometimes be positive, and we shall indeed encounter a few such instances.

In Figures~\ref{fig:pairs xr}--\ref{fig:pairs instruments},
the upper-triangle panels depict the pairs
\[
\bigg({u_i\over \varphi^*(q)}, {\varphi^*(q)v_i \over q^2} \bigg ), \quad i\in \mathcal{M}_q,
\]
whereas the lower-triangle panels depict the pairs
\[
\bigg({u_i\over q}, {v_i\over q} \bigg ), \quad i\in \mathcal{N}_{q},
\]
with specially chosen thresholds $q\in (0,1)$ that we shall specify and discuss later. In every example, we have tested the reasonableness of bound~\eqref{pqd} for paired extreme pseudo-observations $(u_i,v_i)$ using several tests, with findings reported in Appendix~\ref{append-3a}.

\subsection{Foreign currency exchange rates}
\label{appl-1}

\begin{figure}[h!]
    \centering
        \includegraphics[width=.7\textwidth]{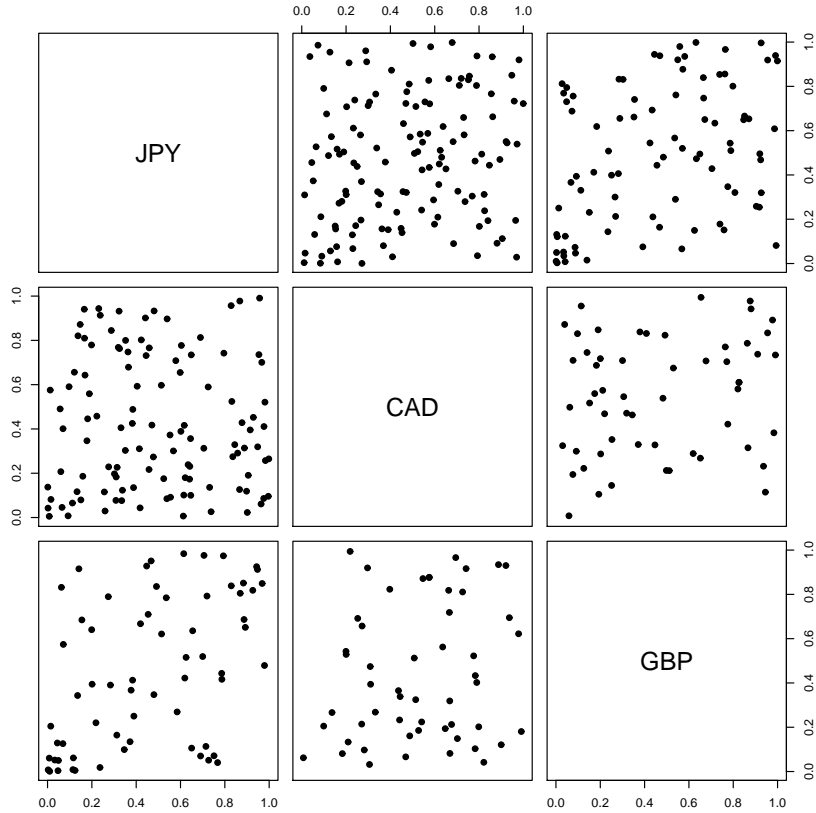}
    \caption{Scatterplots of pseudo observations of foreign currency exchange rates.}
    \label{fig:pairs xr}
\end{figure}

We analyze co-movements of exchange rates of the Canadian and US dollars (CAD/USD), the pound sterling and the US dollar (GBP/USD), and the Japanese yen and the US dollar (JPY/USD) during the period from January 4, 1971, to October 25, 2019 \citep{FRB2020}. The differenced log-exchange rates $(x_t)$ are depicted in the three left-hand panels of Figure~\ref{fig:Log-returns} in Appendix~\ref{append-2}. For typographical simplicity,  we abbreviate CAD/USD, GBP/USD, JPY/USD into CAD, GBP, JPY, respectively. We couple these exchange rates and analyze the strength of their co-movements in regions (determined by $q$) of extreme losses. In Figure~\ref{fig:pairs xr}, %
the thresholds $q = 0.075$, $0.085$, and  $0.1$ have been set for the pairs (JPY, CAD), (JPY, GBP), and (CAD, GBP), respectively.  %
\begin{table}[h!]
\centering
\begin{subtable}{.35\textwidth}\centering
\begin{tabular}{ccc}
\hline\hline
{\cellcolor[gray]{0.8}JPY} & $1.3455$ & $0.9026$ \\
  $123$ & {\cellcolor[gray]{0.8}CAD} & $1.5488$ \\
   $64$ & $57$ & {\cellcolor[gray]{0.8}GBP} \\
    \hline
\end{tabular}
\caption{TOMD $\widehat{\kappa}^*_{m_{q}}(5,\theta_0,q)$ and $m_{q}$.}
\end{subtable}
\qquad
\begin{subtable}{.35\textwidth}\centering
\begin{tabular}{ccc}
\hline\hline
{\cellcolor[gray]{0.8}JPY} & $1.2967$ & $0.8917$ \\
  $112$ & {\cellcolor[gray]{0.8}CAD} & $1.7759$ \\
   $64$ & $53$ & {\cellcolor[gray]{0.8}GBP} \\
    \hline
\end{tabular}
\caption{TODD $\widehat{\kappa}^{\text{OLS}}_{n_q}$ and $n_q$.}
\end{subtable}
\caption{The upper-triangle entries of each panel report estimated tail orders, and the lower-triangle entries report the corresponding sample sizes.}
\label{Tab:results 10per}
\end{table}
Their TOMD and TODD estimates are reported in Table~\ref{Tab:results 10per}.
Note the relative differences:
\begin{align*}
\text{(JPY, CAD)}: \quad
\text{RD} &= 3.76\%
\\
\text{(JPY, GBP)}:  \quad
\text{RD} & =  1.22\%
\\
\text{(CAD, GBP)}:  \quad
\text{RD} & =-12.79 \%
\end{align*}
Although theory says that TOMD is always smaller than TODD, allowing for 5\% variability makes the reported positive percentages unsurprising. We therefore conclude that (JPY, CAD) and (JPY, GBP) must have fairly similar maximal and diagonal tail orders, thus implying strong dependence within the pairs. The remaining third pair (CAD, GBP) shows an almost 13\% relative decrease in the value of TOMD, thus indicating a notable increase in tail dependence when measured by TOMD if compared to TODD.

\subsection{Stock market indices}
\label{appl-2}

\begin{figure}[h!]
    \centering
        \includegraphics[width=.7\textwidth]{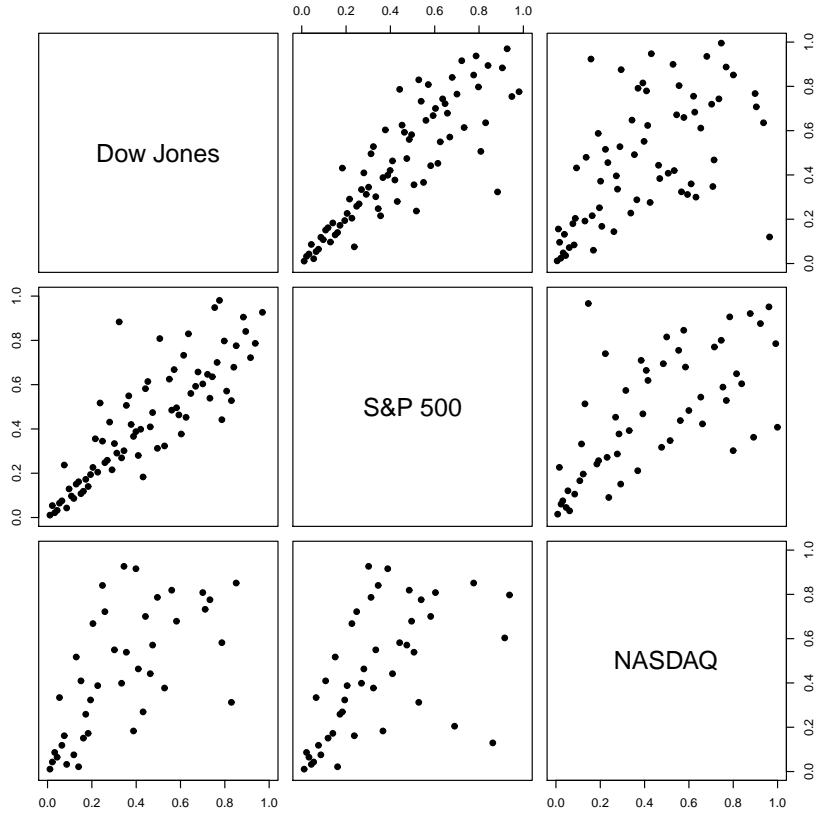}
    \caption{Scatterplots of pseudo observations of stock market indices.}
    \label{fig:pairs indices}
\end{figure}

We analyze extreme co-movements of the Dow Jones, S\&P 500, and NASDAQ during the period from January 4, 1971, to February 28, 2020 \citep{WSJ,Yahoo}. The differenced log-time-series $(x_t)$ are depicted in the three left-hand panels of Figure~\ref{fig:STOCK-Log-returns} in Appendix~\ref{append-2}. We pair these time series and analyze the strength of their co-movements in regions of extreme losses. In Figure~\ref{fig:pairs indices},%
the thresholds $q = 0.0075$, $0.01$, and $0.0075$ have been set for the pairs (Dow Jones, S\&P 500), (Dow Jones, NASDAQ), and (S\&P 500, NASDAQ), respectively. %
\begin{table}[h!]
\centering
\begin{subtable}{.4\textwidth}
\begin{tabular}{ccc}
\hline\hline
{\cellcolor[gray]{0.8}Dow Jones} & $0.8329$ & $0.9816$ \\
 $77$ & {\cellcolor[gray]{0.8}S\&P 500} & $0.9422$ \\
  $68$ & $53$ & {\cellcolor[gray]{0.8}NASDAQ} \\
    \hline
\end{tabular}
\caption{TOMD $\widehat{\kappa}^*_{m_{q}}(5,\theta_0,q)$ and $m_{q}$.}
\end{subtable}
\qquad
\begin{subtable}{.4\textwidth}
\begin{tabular}{ccc}
\hline\hline
{\cellcolor[gray]{0.8}Dow Jones} & $1.0788$ & $0.9874$ \\
 $77$ & {\cellcolor[gray]{0.8}S\&P 500} & $0.9710$ \\
  $61$ & $44$ & {\cellcolor[gray]{0.8}NASDAQ} \\
    \hline
\end{tabular}
\caption{TODD $\widehat{\kappa}^{\text{OLS}}_{n_q}$ and $n_q$.}
\end{subtable}
\caption{The upper-triangle entries of each panel report estimated tail orders, and the lower-triangle entries report the corresponding sample sizes.}
\label{Tab:INDEX-results 10per}
\end{table}
Their TOMD and TODD estimates are reported in Table~\ref{Tab:INDEX-results 10per}.
Note the relative differences:
\begin{align*}
\text{(Dow Jones, S\&P 500)}: \quad
\text{RD} &=  -22.79 \%
\\
\text{(Dow Jones, NASDAQ)}:  \quad
\text{RD} & = -0.59 \%
\\
\text{(S\&P 500, NASDAQ)}:  \quad
\text{RD} & = -2.97 \%
\end{align*}
All the values of TOMD are smaller than the corresponding ones of TODD.
The pair (Dow Jones, S\&P 500) shows an almost 23\% decrease in the value of TOMD if compared to the diagonal case, and thus increase in tail dependence when measured by TOMD.
Allowing for 5\% variability, we conclude that (Dow Jones, NASDAQ) and (S\&P 500, NASDAQ) have quite similar TOMD and TODD, thus implying strong dependence within the pairs.

\subsection{Diverse financial instruments}
\label{appl-3}

\begin{figure}[h!]
    \centering
        \includegraphics[width=.7\textwidth]{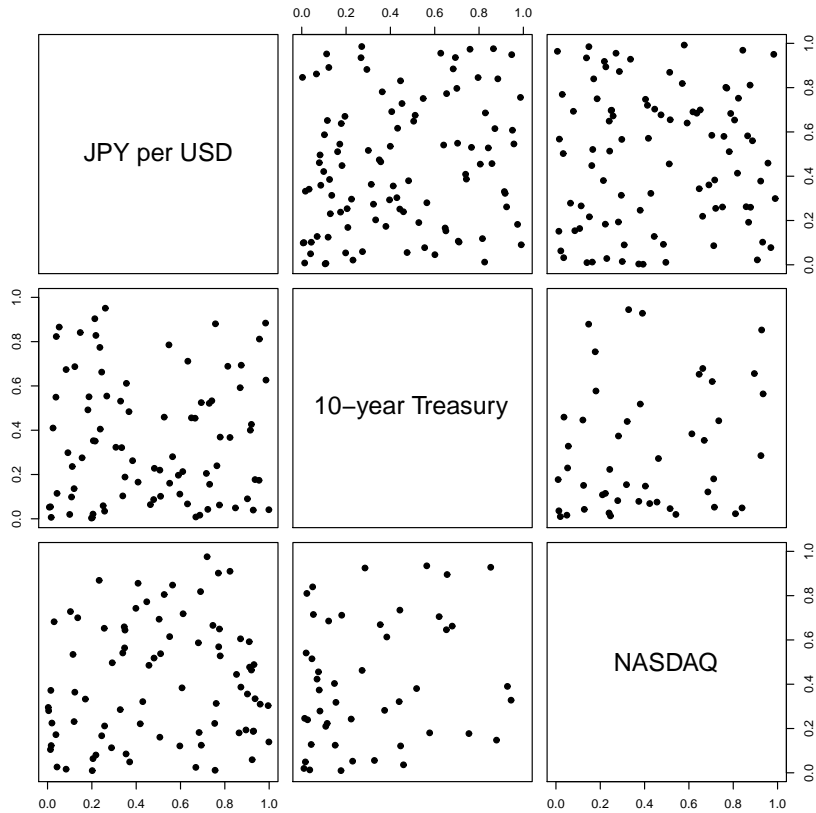}
    \caption{The pseudo observations under consideration of diverse financial instruments.}
    \label{fig:pairs instruments}
\end{figure}

We analyze pairwise extreme co-movements between daily returns of three different financial instruments: JPY/USD \citep{FRB2020}, US 10-year Treasury shorthanded as US10YT, and NASDAQ \citep{Yahoo}, which belong to the categories of exchange rates, treasury notes, and stock market indices, respectively. The historical data are from February 5, 1971, to March 3, 2020.
The differenced log-time-series $(x_t)$ are depicted in the three left-hand panels of Figure~\ref{MIX-fig:Log-returns} in Appendix~\ref{append-2}.
We pair these time series and analyze the strength of their co-movements in regions  of extreme losses. In Figure~\ref{fig:pairs instruments}, %
the thresholds $q = 0.05$, $0.05$, and $0.025$ have been set for the pairs (JPY/USD, US10YT), (JPY/USD, NASDAQ), and (US10YT, NASDAQ), respectively.  %
\begin{table}[h!]
\centering
\begin{subtable}{.4\textwidth}
\begin{tabular}{ccc}
\hline\hline
{\cellcolor[gray]{0.8}JPY/USD} & $1.1648$ & $1.4002$ \\
 $89$ & {\cellcolor[gray]{0.8}US10YT} & $0.8516$ \\
  $87$ & $47$ & {\cellcolor[gray]{0.8}NASDAQ} \\
    \hline
\end{tabular}
\caption{TOMD $\widehat{\kappa}^*_{m_{q}}(5,\theta_0,q)$ and $m_{q}$.}
\end{subtable}
\qquad
\begin{subtable}{.4\textwidth}
\begin{tabular}{ccc}
\hline\hline
{\cellcolor[gray]{0.8}JPY/USD} & $1.3146$ & $1.5276$ \\
 $85$ & {\cellcolor[gray]{0.8}US10YT} & $1.1292$ \\
  $80$ & $47$ & {\cellcolor[gray]{0.8}NASDAQ} \\
    \hline
\end{tabular}
\caption{TODD $\widehat{\kappa}^{\text{OLS}}_{n_q}$ and $n_q$.}
\end{subtable}
\caption{The upper-triangle entries of each panel report estimated tail orders, and the lower-triangle entries report the corresponding sample sizes.}
\label{MIX-Tab:results 10per}
\end{table}
Their TOMD and TODD estimates are reported in Table~\ref{MIX-Tab:results 10per}.
Note the relative differences:
\begin{align*}
\text{(JPY/USD, US10YT)}: \quad
\text{RD} &= -11.40 \%
\\
\text{(JPY/USD, NASDAQ)}:  \quad
\text{RD} & = -8.34 \%
\\
\text{(US10YT, NASDAQ)}:  \quad
\text{RD} & = -24.58 \%
\end{align*}
All the values of TOMD are smaller than the corresponding ones of TODD, and all the pairs show considerable decrease in the values of TOMD if compared to the diagonal case.

\section{Conclusion}\label{sec: conclusion}

In this paper we have developed a substantial extension of the procedure of \citet{SYZ2020} for assessing the maximal strength of co-movements of extreme losses when original data follow dependent dynamical models with the underlying copulas whose maximal tail probabilities near the origin are only known to be regularly varying functions.  We have explored the performance of the modification on simulated bivariate time series. Our study has shown that the block-wise construction of the estimator of maximal tail dependence successfully handles time series structures and from them arising extreme co-movements.

We have tested the validity of underlying theoretical assumptions using several statistical tests, and also discussed ways for calculating their critical values. In addition, we have provided an extensive study of thresholds below which time-series data give rise to extremes.

The strength of maximal dependence as well as of the classical diagonal dependence have been explored and compared for a number of financial instruments,  such as foreign exchange rates of several major currencies, stock market indices, and treasury notes.

\section*{Acknowledgements}

We are grateful to the anonymous reviewers for their expert analysis of our technical and numerical results, constructive criticism, suggestions and insights, all of which helped us to prepare a much improved version of the paper. Thanks are due to Takaaki Koike for his illuminating presentation at the Risk Management and Actuarial Science Seminar (University of Waterloo and Tsinghua University) on the topic of measuring tail dependence, and also for his insights shared with us thereafter.

\section*{Funding}

This research has been supported by the Natural Sciences and Engineering Research Council (NSERC) of Canada, and the national research organization Mathematics of Information Technology and Complex Systems (MITACS) of Canada.

\section{Conflict of interest}

The authors declare that they have no conflicts of interest.


\appendix

\section{Technicalities}
\label{append-00}

The proof of Theorem~\ref{thm: essinf} is long, but it is necessary to present in order to see why and how the estimator works. We note at the outset that the joint cdf $F_q^*$ may not have the uniform on $[0,1]$ marginal distributions, and so bound~\eqref{pqd} does not really mean that $F_q^*$ is PQD. Nevertheless, it is this bound that we need in the proof of Theorem~\ref{thm: essinf}. In the following appendices we shall test the validity of this bound using a number of statistical tests.

\subsection{Proof of Theorem~\ref{thm: essinf}}
\label{proof-of-th31}

We start the proof by expressing $\kappa^*(m,0,q) $ in terms of the survival function of the random variable
\begin{equation}\label{xi-def}
\xi_q:={2\log (C(U_q,V_q)/ \Pi^*(q))\over \log (U_q V_q/ q^2) },
\end{equation}
where $(U_q,V_q)$ follows the cdf $F_q$.
That is, we have the equation
\begin{equation}\label{new-eq-0}
\kappa^*(m,0,q) =\int_0^{\infty}\mathbb{P}(\xi_q>x)^m\dif x
\end{equation}
because the cardinality of the set $G_{1,q}$ is $m$ and the pairs $(U_{i,q},V_{i,q})$, $i\in G_{1,q}$, are iid. With the notation
\[
(\widetilde{U}_q,\widetilde{V}_q):=\bigg ( {U_q\over \varphi^*(q)},
{V_q  \varphi^*(q)\over q^2}\bigg ),
\]
we rewrite $\xi_q$ as
\begin{align}
\xi_q
&={2\log (C(U_q,V_q)/\Pi^*(q))\over \log (U_q/\varphi^*(q))+\log(V_q\varphi^*(q)/ q^2)}
\notag
\\
&= {2\log F_q^*(\widetilde{U}_q,\widetilde{V}_q)\over \log \widetilde{U}_q+\log\widetilde{V}_q} \in [0,2],
\label{xi-def-1}
\end{align}
where the inclusion into the interval $[0,2]$ is due to the assumed bound~\eqref{pqd}.
Hence, bound~\eqref{new-eq-0} reduces to
\begin{equation}\label{eq-99a}
\kappa^*(m,0,q) = \int_0^{2}\mathbb{P}(\xi_q>x)^m\dif x  .
\end{equation}

\begin{note}
The intuitive meaning of equation~\eqref{xi-def-1} is to scale the pair $(U_{q},V_{q})\in \mathcal{R}_q(\mathbf{0})$ into $(\widetilde{U}_{q},\widetilde{V}_{q})\in [0,1]\times [0,1]$.  This allows us to shift the focus from the behaviour of random pairs with respect to $q\downarrow 0$ toward the behaviour of $F_q^*$ and the scaling parameters $\varphi^*(q)$ and $q^2/\varphi^*(q)$.
\end{note}

Since $(U_{q},V_{q})\in \mathcal{R}_q(\mathbf{0})$ and the functions $\varphi^*$ and $z\mapsto  z^2/\varphi^*(z)$ are increasing, we have
\[
(\widetilde{U}_{q}^*,\widetilde{V}_{q}^*):=\bigg ({\varphi^*(\sqrt{U_{q}V_{q}})\over \varphi^*(q)} ,{(\sqrt{U_{q}V_{q}})^2/\varphi^*(\sqrt{U_{q}V_{q}}) \over q^2/\varphi^*(q)} \bigg ) \in [0,1]\times [0,1].
\]
Hence,
\begin{align*}
F_q^*(\widetilde{U}_{q}^*,\widetilde{V}_{q}^*)
&= {1\over \Pi^*(q)} C\big(\varphi^*(\sqrt{U_{q}V_{q}}), U_{q}V_{q}/\varphi^*(\sqrt{U_{q}V_{q}})\big)
\\
&={1\over \Pi^*(q)} \sup_{x\in [U_{q}V_{q},1]} C\big(x, U_{q}V_{q}/x)\big)
\\
&\ge {1\over \Pi^*(q)}C(U_q,V_q)
\\
&= F_q^*(\widetilde{U}_q,\widetilde{V}_q).
\end{align*}
Since
$\widetilde{U}_{q}^*\widetilde{V}_{q}^* = \widetilde{U}_q\widetilde{V}_q = U_{q}V_{q}/q^2\in(0,1)$,
we arrive at the bound
\begin{equation}\label{eq-string-01}
\dfrac{2\log F_q^*(\widetilde{U}_{q}^*,\widetilde{V}_{q}^*)}{\log\widetilde{U}_{q}^* + \log\widetilde{V}_{q}^*} \le \dfrac{2\log F_q^*(\widetilde{U}_q,\widetilde{V}_q)}{\log\widetilde{U}_q + \log\widetilde{V}_q}.
\end{equation}
Note that
\begin{align}\label{eq-string-0}
{2\log F_q^*(\widetilde{U}_{q}^*,\widetilde{V}_{q}^*)\over \log \widetilde{U}_{q}^*+\log\widetilde{V}_{q}^*}
&= {2\log \Big(C\big(\varphi^*(\sqrt{U_{q}V_{q}}),
(\sqrt{U_{q}V_{q}})^2/\varphi^*(\sqrt{U_{q}V_{q}})\big)/ \Pi^*(q)\Big)\over \log (U_{q}V_{q}/ q^2)}
\notag
\\
&= {2\log \big(\Pi^*(\sqrt{U_{q}V_{q}})/ \Pi^*(q)\big)\over \log (U_{q}V_{q}/ q^2)}.
\end{align}
Due to equation~\eqref{max-tail eq}, we have
\[
{\Pi^*(\sqrt{U_{q}V_{q}})\over \Pi^*(q)}
= {\ell^*(\sqrt{U_{q}V_{q}}) \over \ell^*(q)}
\bigg({U_{q}V_{q}\over q^2}\bigg)^{\kappa^*/2}
\]
and thus, continuing with equation~\eqref{eq-string-0} and taking into account bound~\eqref{eq-string-01}, we obtain
\begin{equation}\label{eq-string-1}
\dfrac{2\log F_q^*(\widetilde{U}_q,\widetilde{V}_q)}{\log\widetilde{U}_q + \log\widetilde{V}_q}
\ge {2\log (\ell^*(\sqrt{U_{q}V_{q}}) / \ell^*(q)) \over \log (U_{q}V_{q}/ q^2)}
+ \kappa^*  .
\end{equation}
With the notation
\[
o_q := {2\log (\ell^*(\sqrt{U_{q}V_{q}}) / \ell^*(q)) \over \log (U_{q}V_{q}/ q^2)},
\]
bound~\eqref{eq-string-1} takes the form
\begin{equation}\label{eq-string-ineq}
\xi_q \ge \max\{0,o_q + \kappa^*  \}.
\end{equation}

We next prove the statement
\begin{equation}\label{eq-string-3}
o_q \stackrel{\mathbb{P}}{\to} 0
\end{equation}
when $q\downarrow 0$. With the notation $W_q=\sqrt{U_{q}V_{q}}/q$, statement~\eqref{eq-string-3} is equivalent to
\[
o_q ={\log (\ell^*(qW_q) / \ell^*(q))\over \log W_q}\stackrel{\mathbb{P}}{\to} 0 .
\]
To prove it, we fix $\varepsilon>0$ and $\delta>0$, with the latter parameter used to partition the sample space into the following three events: $\{W_q< \delta\}$, $\{\delta \le W_q\le 1-\delta\}$ and $\{W_q> 1-\delta\}$. We obtain the bound
\begin{multline}\label{eq-string-5a}
\mathbb{P}\bigg(\bigg|{\log (\ell^*(qW_q) / \ell^*(q))\over \log W_q}\bigg|>\varepsilon\bigg)
\\
\le \mathbb{P}\bigg(\sup_{w\in [\delta,1-\delta]}\bigg|{\log (\ell^*(qw) / \ell^*(q))\over \log w}\bigg| > \varepsilon\bigg)
+ \mathbb{P}\left(W_q<\delta\right)
+ \mathbb{P}\left(W_q>1-\delta\right).
\end{multline}
Since $\ell^*$ is slowly varying at $0$, we have
$\ell^*(qw)/ \ell^*(q)\to 1$ when $ q \downarrow 0$ for every $w\in (0,1]$. The convergence is uniform in $w\in [\delta,1]$ for any fixed $\delta>0$, which implies \citep[Lemma 1, p.~310]{BS1971} that $\sup_{w\in [\delta,1]}|\log (\ell^*(qw) / \ell^*(q))|$ converges to $0$ when $ q \downarrow 0$. Hence, we conclude from bound~\eqref{eq-string-5a} that, for any $\varepsilon>0$ and $\delta>0$,
\begin{equation}\label{eq-string-5}
\limsup_{q\to 0}\mathbb{P}\bigg(\bigg|{\log (\ell^*(qW_q) / \ell^*(q))\over \log W_q}\bigg|>\varepsilon\bigg)
\\
\le \limsup_{q\to 0} \mathbb{P}\left(W_q<\delta\right)
+ \limsup_{q\to 0}\mathbb{P}\left(W_q>1-\delta\right).
\end{equation}
Note that the left-hand side of bound~\eqref{eq-string-5} does not depend on $\delta >0$. As to the first probability on the right-hand side of bound~\eqref{eq-string-5}, we have
\begin{align}\label{eq-w9}
\mathbb{P}(W_q \le \delta)
&= \mathbb{P}(U_qV_q \le \delta\varphi^*(q) \delta q^2/\varphi^*(q) )
\notag
\\
&\le \mathbb{P}(U_q\le \delta\varphi^*(q)) + \mathbb{P}(V_q\le \delta q^2/\varphi^*(q)) \notag
\\
&= \dfrac{1}{\Pi^*(q)}\left(C(\delta\varphi^*(q), q^2/\varphi^*(q)) + C(\varphi^*(q), \delta q^2/\varphi^*(q))\right)
\notag
\\
&\le \dfrac{2\Pi^*(q\sqrt{\delta})}{\Pi^*(q)}
\notag
\\
&\to 2\delta^{\kappa^*/2}
\end{align}
when $q\downarrow 0$, where we used equation~\eqref{max-tail eq} and property~\eqref{ratio-ells}.

We tackle the second probability on the right-hand side of bound~\eqref{eq-string-5} in a different way, starting as follows:
\begin{align}
\mathbb{P}(W_q > 1-\delta)
&= \mathbb{P}(\widetilde{U}_q\widetilde{V}_q > (1-\delta)^2)
\notag
\\
&\le \mathbb{P}(F^*_q(\widetilde{U}_q,\widetilde{V}_q) > (1-\delta)^2),
\label{eq-w10}
\end{align}
where the bound holds because $F^*_q(u,v) \ge uv$ for all $u,v\in [0,1]$. With the notation
$$
K^*_q(t) := \mathbb{P}(F^*_q(\widetilde{U}_q,\widetilde{V}_q)\le t)
$$
we continue with bound~\eqref{eq-w10} and have
\begin{align}
\mathbb{P}(W_q > 1-\delta)
&= 1 - K^*_q((1-\delta)^2)
\notag
\\
&\le 1 - (1 - \delta)^2
\label{eq-w11}
\end{align}
for every $q\in (0,1]$, because
\begin{align*}
K^*_q(t)
&\ge \mathbb{P}(F^*_q(\widetilde{U}_q,1)\le t)
\\
&= \mathbb{P}(G_q(\widetilde{U}_q)\le t)
\\
&= \mathbb{P}(\widetilde{U}_q\le G^{-1}_q(t))
\\
&= G_q\circ G^{-1}_q(t)
\\
&= t,
\end{align*}
where $G_q:[0,1]\to [0,1]$ denotes the cdf of $\widetilde{U}_q$ given by
\begin{align*}
G_q(u)
&= \dfrac{C(\varphi^*(q)u,q^2/\varphi^*(q))}{C(\varphi^*(q),q^2/\varphi^*(q))}
\\
&= \dfrac{C(\varphi^*(q)u,q^2/\varphi^*(q))}{\Pi^*(q)}.
\end{align*}
Hence, in view of bounds~\eqref{eq-w9} and \eqref{eq-w11}, the entire right-hand side of bound~\eqref{eq-string-5} vanishes when $\delta \downarrow 0$. This concludes the proof of statement~\eqref{eq-string-3}.

Fix now any $\varepsilon \in (0,\kappa^*)$, which, by the way, has nothing to do with the earlier used $\varepsilon $. Equation~\eqref{eq-99a} implies the bound
\begin{equation}\label{eq-99b}
\kappa^*(m,0,q) \le \kappa^* + \varepsilon + \int_{\kappa^*+\varepsilon }^{2}\mathbb{P}(\xi_q>x)^m\dif x .
\end{equation}
To estimate $\kappa^*(m,0,q)$ from below, we start with
\begin{align}
\kappa^*(m,0,q)
&= \kappa^* - \varepsilon
- \int_0^{\kappa^*-\varepsilon }1-\mathbb{P}(\xi_q>x)^m\dif x
+ \int_{\kappa^*+\varepsilon }^{2}\mathbb{P}(\xi_q>x)^m\dif x
\notag
\\
&\ge \kappa^* - \varepsilon
- (\kappa^*-\varepsilon)m\mathbb{P}(\xi_q\le \kappa^*-\varepsilon)
+ \int_{\kappa^*+\varepsilon }^{2}\mathbb{P}(\xi_q>x)^m\dif x .
\label{eq-u1}
\end{align}
Using bound~\eqref{eq-string-ineq}, we obtain
\begin{align}
\mathbb{P}(\xi_q\le \kappa^*-\varepsilon)
&\le \mathbb{P}\big (\max\{0,o_q + \kappa^*  \} \le \kappa^*-\varepsilon \big )
\notag
\\
&= \mathbb{P}\big (o_q  \le -\varepsilon, o_q + \kappa^*> 0 \big )
+\mathbb{P}\big (o_q + \kappa^*\le 0\big )
\notag
\\
&= 2 \mathbb{P}(|o_q | \ge \varepsilon ).
\label{eq-u2}
\end{align}
The right-hand side can be made as small as desired by choosing a sufficiently small $q>0$. Hence, due to  bounds~\eqref{eq-99b}--\eqref{eq-u2}, for any $m\ge 1$ we can choose sufficiently small $\varepsilon >0$ and $q>0$ such that $\kappa^*(m,0,q) $ is as close to $\kappa^* $ as desired, provided that the integral $\int_{\kappa^*}^{2}\mathbb{P}(\xi_q>x)^m\dif x $ can be made as small as desired by choosing a sufficiently large $m$.

To prove the latter statement, without loss of generality we assume $\kappa^*<2$, which prevents $F_q^*$ from being the independence copula. By the Lebesgue dominated convergence theorem, the integral $\int_{\kappa^*}^{2}\mathbb{P}(\xi_q>x)^m\dif x $ converges to $0$ when $m\to \infty $ provided that $\mathbb{P}(\xi_q>x)<1$ for all $x\in (\kappa^*,2)$. Hence, we need to show that
\begin{equation}\label{eq-99c}
\mathbb{P}(\xi_q  \le \kappa^*+h ) > 0 \quad \textrm{for every} \quad h\in(0, 2-\kappa^*).
\end{equation}
Although the proof of this statement follows the ideas of \citet[Theorem 3.2]{SYZ2020}, substantial adjustments are required, which we give next.

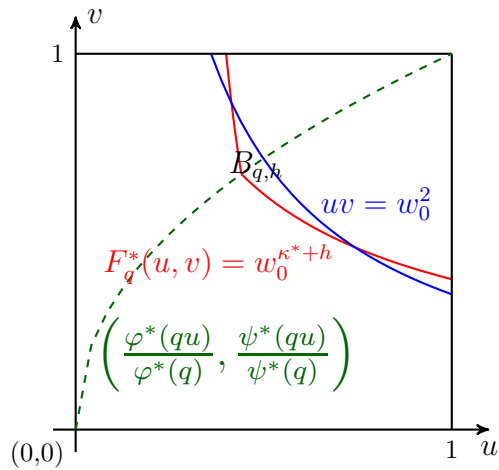
\begin{figure}[h!]
\centering
\begin{tikzpicture}
[
    thick,
    >=stealth',
    dot/.style = {
      draw,
      fill = white,
      circle,
      inner sep = 0pt,
      minimum size = 4pt
    }
  ]
  \coordinate (O) at (0,0);
  \draw[->] (-0.3,0) -- (5.5,0) coordinate[label = {below:$u$}] (xmax);
  \draw[->] (0,-0.3) -- (0,5.5) coordinate[label = {right:$v$}] (ymax);
  \draw [red] plot[domain=2.202491:5] (\x, {(\x / 5)^(0.3529 - 1)*2});
  \draw [red] plot[domain=3.399647:5] ({(\x / 5)^(0.75 - 1)*2},\x);
  \draw (5,5) -- (0,5) node[pos=1, left] {\footnotesize 1};
  \draw (5,5) -- (5,0) node[pos=1, below] {\footnotesize 1};
  \node  [pos=1, below left] {\footnotesize (0,0)};
  \draw [green!40!black, dashed] plot[domain=0:5] (\x, {(\x/5)^(0.3529 / 0.75)*5});
  \node[green!40!black] at (2.0,1.0) {\scalebox{1.3}{$\left({\varphi^*(qu)\over \varphi^*(q)}, {\psi^*(qu)\over \psi^*(q)}\right)$}};
  \draw [blue] plot[domain=1.8:5] (\x, {9 / \x});
  \node at (2.4,3.5) {$B_{q,h}$};
  \node[red] at (1.9,2.2) {$F^*_q(u,v) = w_0^{\kappa^*+h}$};
  \node[blue] at (4.0,3.0) {$uv = w^2_0$};
\end{tikzpicture}
\caption{The set $B_{q,h}$ and associated curves.}
\label{figure-0}
\end{figure}

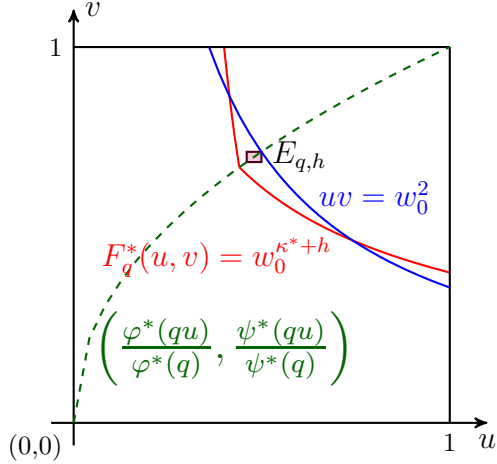
\begin{figure}[h!]
\centering
\begin{tikzpicture}
[
    thick,
    >=stealth',
    dot/.style = {
      draw,
      fill = white,
      circle,
      inner sep = 0pt,
      minimum size = 4pt
    }
  ]
  \coordinate (O) at (0,0);
  \draw[->] (-0.3,0) -- (5.5,0) coordinate[label = {below:$u$}] (xmax);
  \draw[->] (0,-0.3) -- (0,5.5) coordinate[label = {right:$v$}] (ymax);
  \filldraw[fill=purple!20!white, draw=purple!40!black] (2.300327,3.469887) rectangle (2.495999,3.60577);
  \draw [red] plot[domain=2.202491:5] (\x, {(\x / 5)^(0.3529 - 1)*2});
  \draw [red] plot[domain=3.399647:5] ({(\x / 5)^(0.75 - 1)*2},\x);
  \draw (5,5) -- (0,5) node[pos=1, left] {\footnotesize 1};
  \draw (5,5) -- (5,0) node[pos=1, below] {\footnotesize 1};
  \node  [pos=1, below left] {\footnotesize (0,0)};
  \draw [blue] plot[domain=1.8:5] (\x, {9 / \x});
  \draw [green!40!black, dashed] plot[domain=0:5] (\x, {(\x/5)^(0.3529 / 0.75)*5});
  \node[green!40!black] at (2.0,1.0) {\scalebox{1.3}{$\left({\varphi^*(qu)\over \varphi^*(q)}, {\psi^*(qu)\over \psi^*(q)}\right)$}};
  \node[left, below] at (3.0,3.9) {$E_{q,h}$};
  \node[red] at (1.9,2.2) {$F^*_q(u,v) = w_0^{\kappa^*+h}$};
  \node[blue] at (4.0,3.0) {$uv = w^2_0$};
\end{tikzpicture}
\caption{The rectangle $E_{q,h}\subset B_{q,h}$ and associated curves.}
\label{figure-1}
\end{figure}

We start with the bound
\begin{align}
\mathbb{P}(\xi_q \le \kappa^*+h )
&=\mathbb{P}\bigg(\dfrac{2\log F_q^*(\widetilde{U}_q, \widetilde{V}_q)}{\log \widetilde{U}_q + \log \widetilde{V}_q}\le \dfrac{2\log w^{\kappa^*+h}_0}{\log w^2_0}\bigg)
\notag
\\
&\ge \mathbb{P}\left((\widetilde{U}_q, \widetilde{V}_q)\in B_{q,h} \right),
\label{eq-99d}
\end{align}
where, for some $w_0\in(0,1)$,
$$
B_{q,h} := \big \{(u,v)\in[0,1]\times [0,1]: uv \le w^2_0, ~ F_q^*(u,v) > w^{\kappa^*+h}_0 \big \},
$$
which is depicted in Figure~\ref{figure-0}.
With the notation $w_0=\sqrt{u_qv_q}$, we have
\[
u_q = \varphi^*(qw_0)/\varphi^*(q), \quad v_q = w^2_0\varphi^*(q)/\varphi^*(qw_0),
\]
and
\begin{align*}
F_q^*(u_q, v_q)
&= \dfrac{\ell^*(qw_0)}{\ell^*(q)}w^{\kappa^*}_0
\\
&\approx w^{\kappa^*}_0
\\
&>w^{\kappa^*+h}_0.
\end{align*}
Hence,
$F_q^*(u_q, v_q) > w^{\kappa^*+h}_0$
for a sufficiently small $q>0$ (depending on $w_0$ and $h$). Note that
\begin{align*}
F_q^*\bigg(\dfrac{\varphi^*(qw^{1+h/2\kappa^*}_0)}{\varphi^*(q)}, \dfrac{w^{2+h/\kappa^*}_0\varphi^*(q)}{\varphi^*(qw^{1+h/2\kappa^*}_0)}\bigg)
&=\dfrac{\Pi^*(qw^{1+h/2\kappa^*}_0)}{\Pi^*(q)}
\\
&\approx w^{\kappa^*+h/2}_0.
\end{align*}
We have
$$
F_q^*\bigg(\dfrac{\varphi^*(qw^{1+h/2\kappa^*}_0)}{\varphi^*(q)}, \dfrac{w^{2+h/\kappa^*}_0\varphi^*(q)}{\varphi^*(qw^{1+h/2\kappa^*}_0)}\bigg)
> w^{\kappa^*+h}_0
$$
for as sufficiently small $q>0$. With
$$
u_{q,h} := \dfrac{\varphi^*(qw^{1+h/2\kappa^*}_0)}{\varphi^*(q)}
\quad \textrm{and} \quad
v_{q,h} := \dfrac{w^{2+h/\kappa^*}_0\varphi^*(q)}{\varphi^*(qw^{1+h/2\kappa^*}_0)},
$$
we have
\[
u_{q,h}v_{q,h} = w^{2+h/\kappa^*}_0 < w^2_0.
\]
Thus, $(u_{q,h}, v_{q,h})\in B_{q,h}$. In particular, $(u_{q,0}, v_{q,0}) = (u_q, v_q)$. Since $\varphi^*$ and $\psi^*$ are strictly increasing functions, we have $u_{q,h} < u_q$ and $v_{q,h} < v_q$, and so the rectangle
\[
E_{q,h} := (u_{q,h}, u_q]\times(v_{q,h}, v_q],
\]
which we depict in Figure~\ref{figure-1},
is a non-empty subset of $B_{q,h}$.
We have
\begin{align*}
\mathbb{P}\big((\widetilde{U}_q, \widetilde{V}_q) \in E_{q,h}\big)
&= F_q^*(u_q,v_q) - F_q^*(u_q, v_{q,h}) - F_q^*(u_{q,h}, v_q) + F_q^*(u_{q,h},v_{q,h})
\\
&= \dfrac{1}{\Pi^*(q)}\left (
C(\varphi^*(qw_0), \psi^*(qw_0)) - C(\varphi^*(qw_0), \psi^*(qw^{1+h/2\kappa^*}_0) \right.
\\
&\qquad \qquad \left.- C(\varphi^*(qw^{1+h/2\kappa^*}_0), \psi^*(qw_0)) + C(\varphi^*(qw^{1+h/2\kappa^*}_0), \psi^*(qw^{1+h/2\kappa^*}_0))
\right)
\\
&\ge \dfrac{w^{\kappa^*}_0}{\ell^*(q)}\left(
\ell^*(qw_0) - 2\ell^*(qw^{1+h/4\kappa^*}_0)w^{h/4}_0 + \ell^*(qw^{1+h/2\kappa^*}_0)w^{h/2}_0
\right)
\\
&\to w^{\kappa^*}_0(1 - w^{h/4}_0)^2 > 0
\end{align*}
when $q\downarrow 0$. In summary, we have proved $\mathbb{P}((\widetilde{U}_q, \widetilde{V}_q) \in E_{q,h}) > 0$ for a sufficiently small $q>0$. This completes the proof of Theorem~\ref{thm: essinf}.

\subsection{Thresholds and pseudo observations}
\label{append-2}

\begin{figure}[h!]
\centering
    \begin{subfigure}[b]{0.48\textwidth}
        \includegraphics[width=\textwidth]{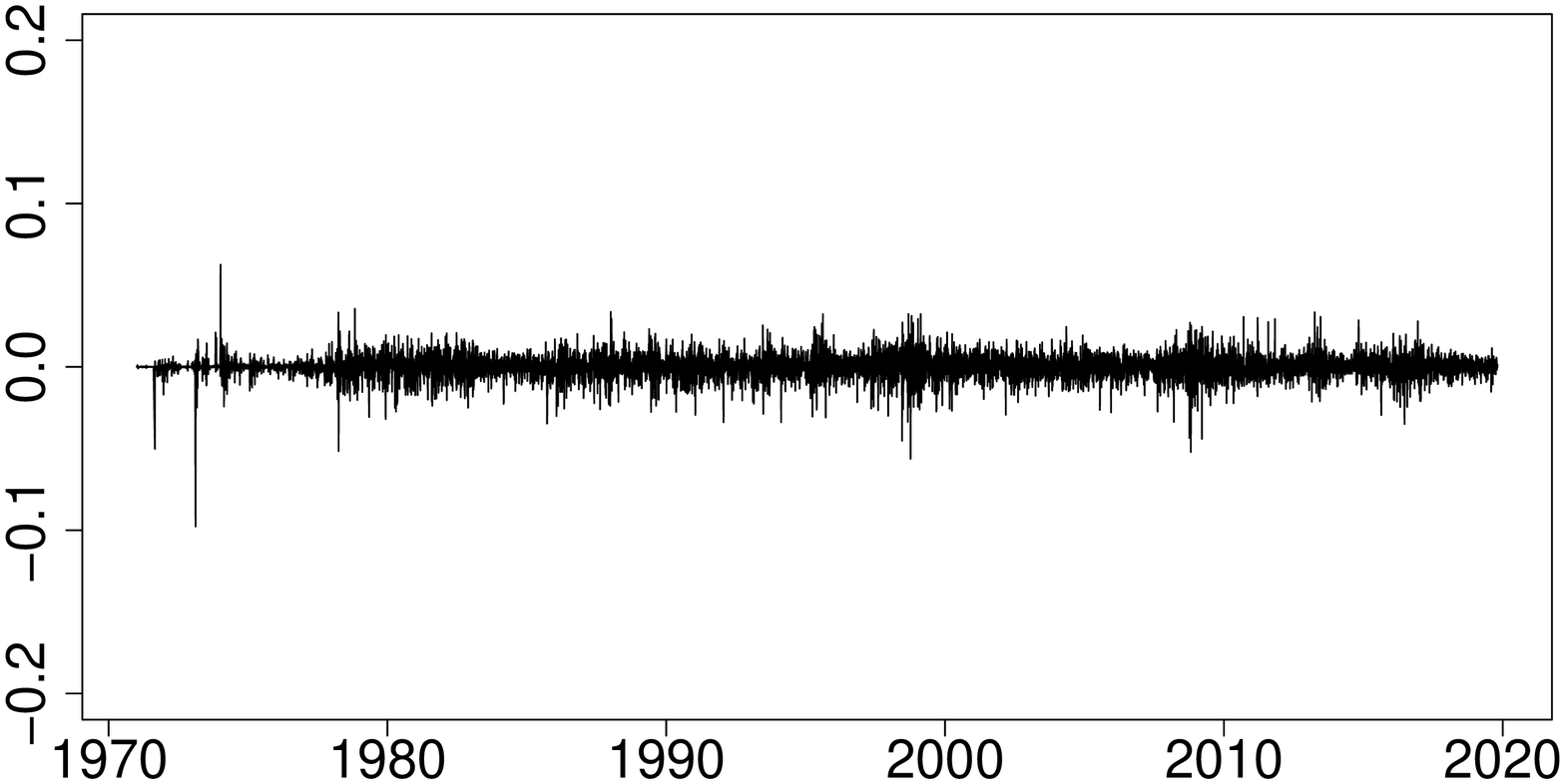}
        \caption{JPY/USD}
        \label{JPY/USD}
    \end{subfigure}
\hfill
    \begin{subfigure}[b]{0.48\textwidth}
        \includegraphics[width=\textwidth]{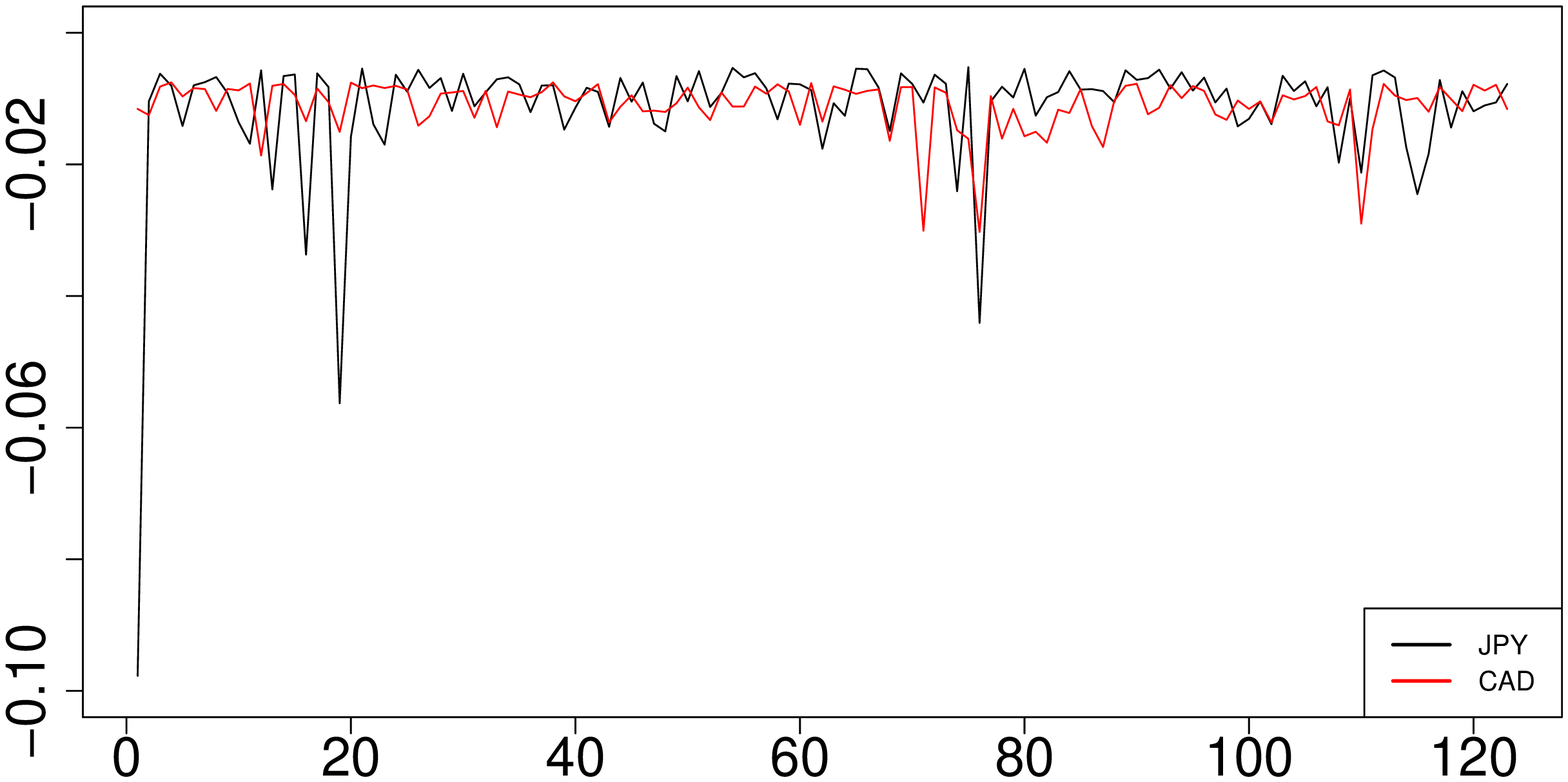}
        \caption{(JPY, CAD), $q = 0.075$, $m_{q} = 123$.}
        \label{Fig: JPY vs. CAD}
    \end{subfigure}
\\
    \begin{subfigure}[b]{0.48\textwidth}
        \includegraphics[width=\textwidth]{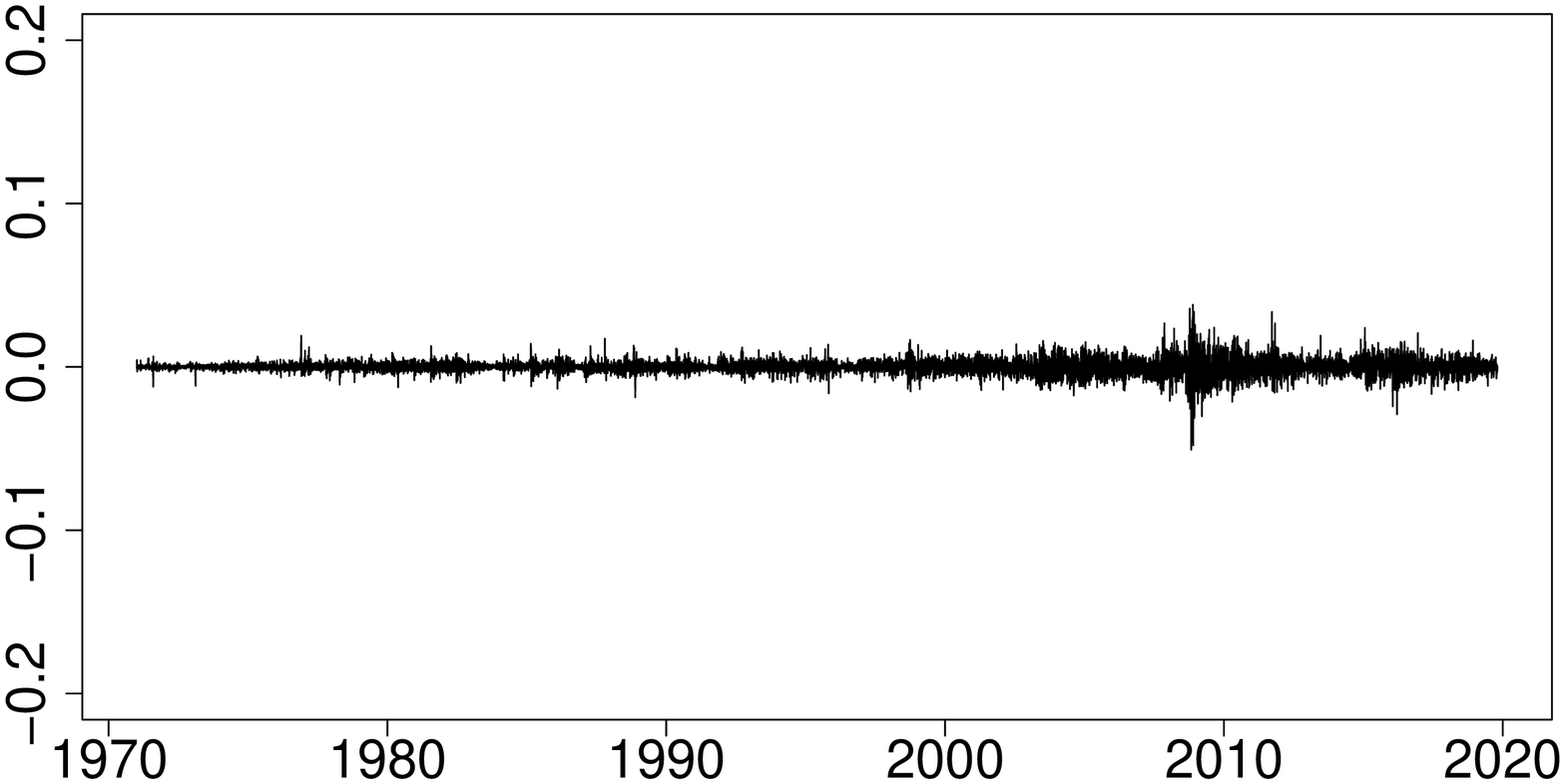}
        \caption{CAD/USD}
        \label{CAD/USD}
    \end{subfigure}
\hfill
    \begin{subfigure}[b]{0.48\textwidth}
        \includegraphics[width=\textwidth]{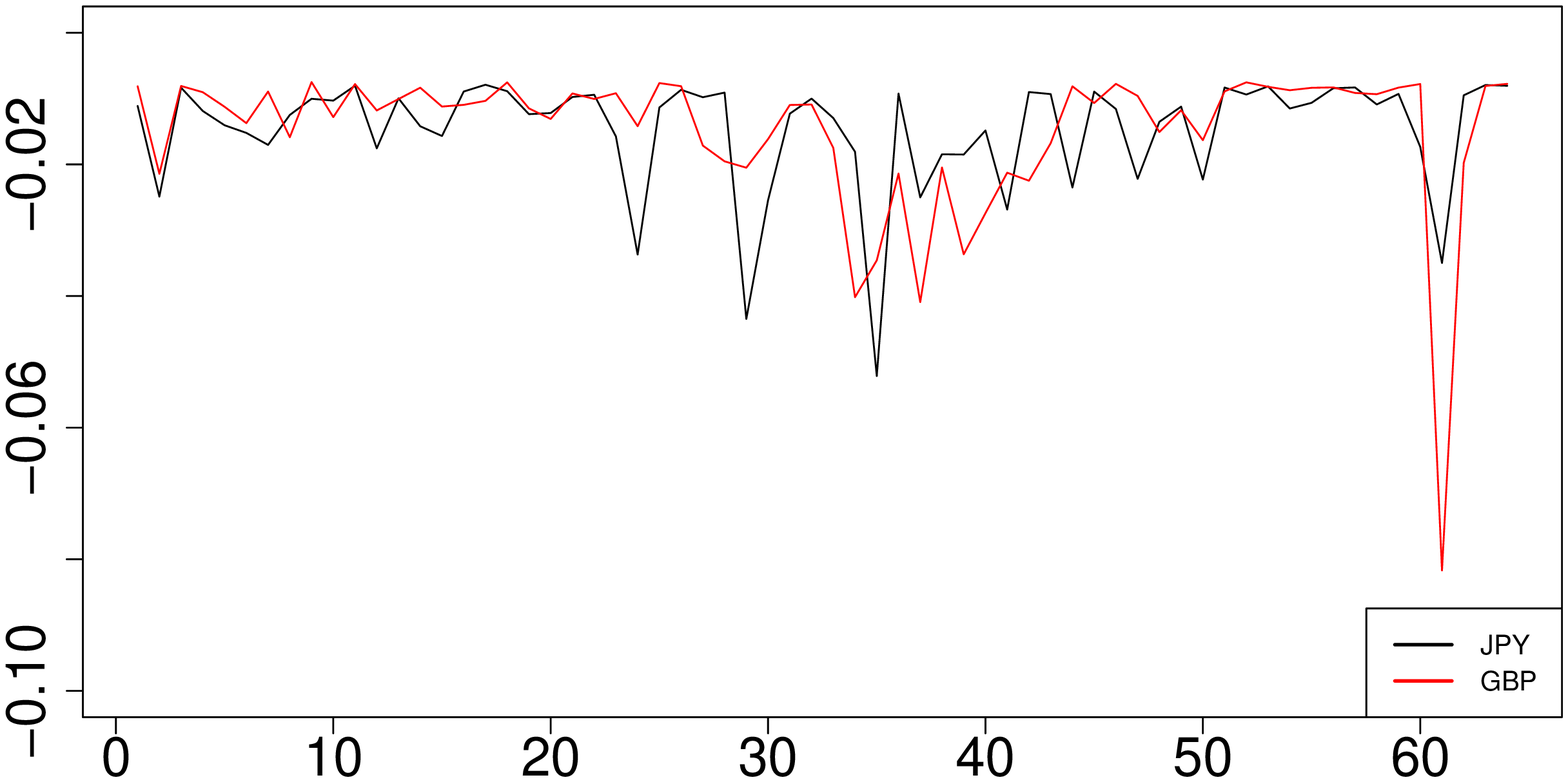}
        \caption{(JPY, GBP), $q = 0.085$, $m_{q} = 64$.}
        \label{Fig: JPY vs. GBP}
    \end{subfigure}
\\
    \begin{subfigure}[b]{0.48\textwidth}
        \includegraphics[width=\textwidth]{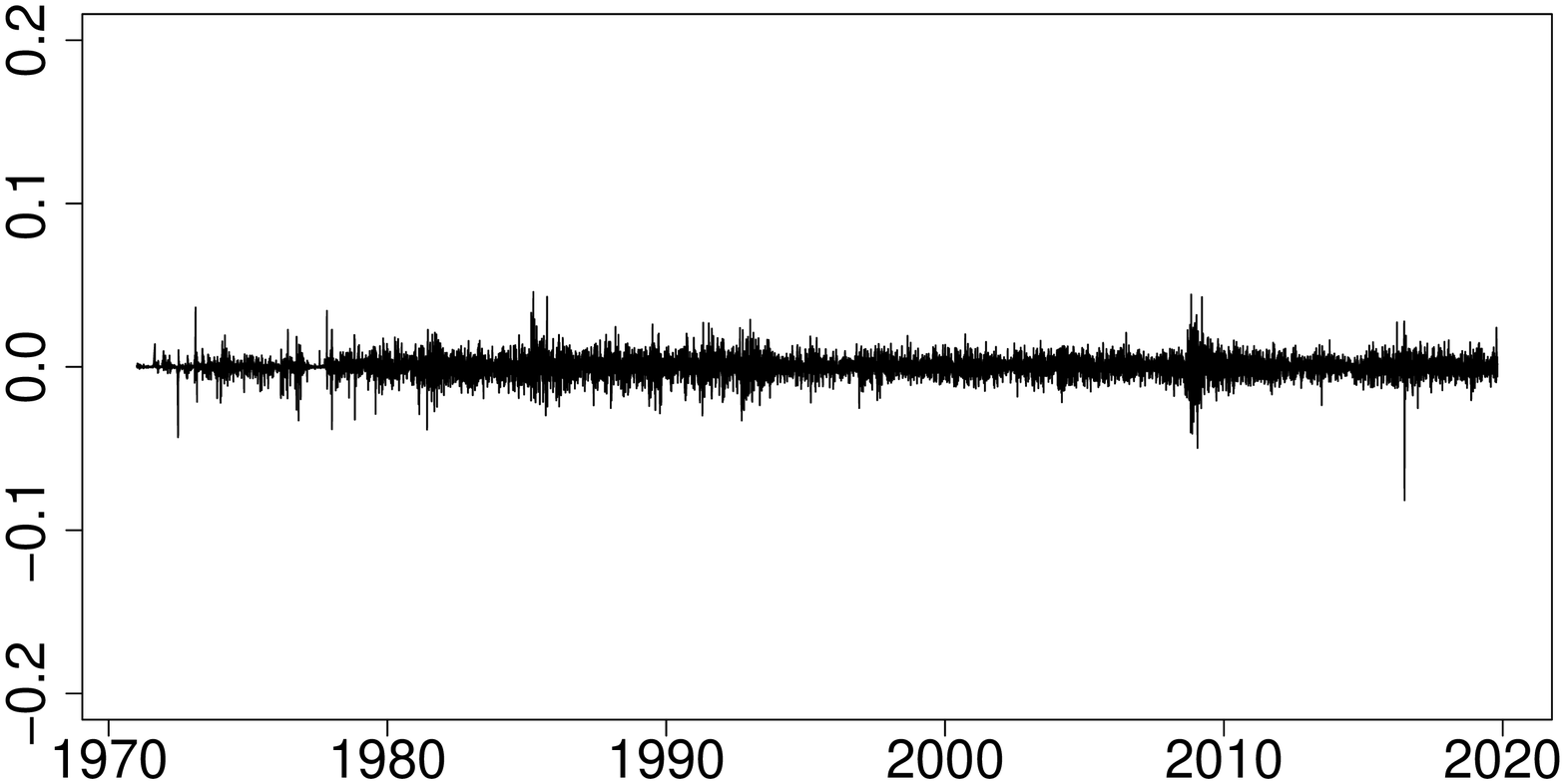}
        \caption{GBP/USD}
        \label{GBP/USD}
    \end{subfigure}
\hfill
    \begin{subfigure}[b]{0.48\textwidth}
        \includegraphics[width=\textwidth]{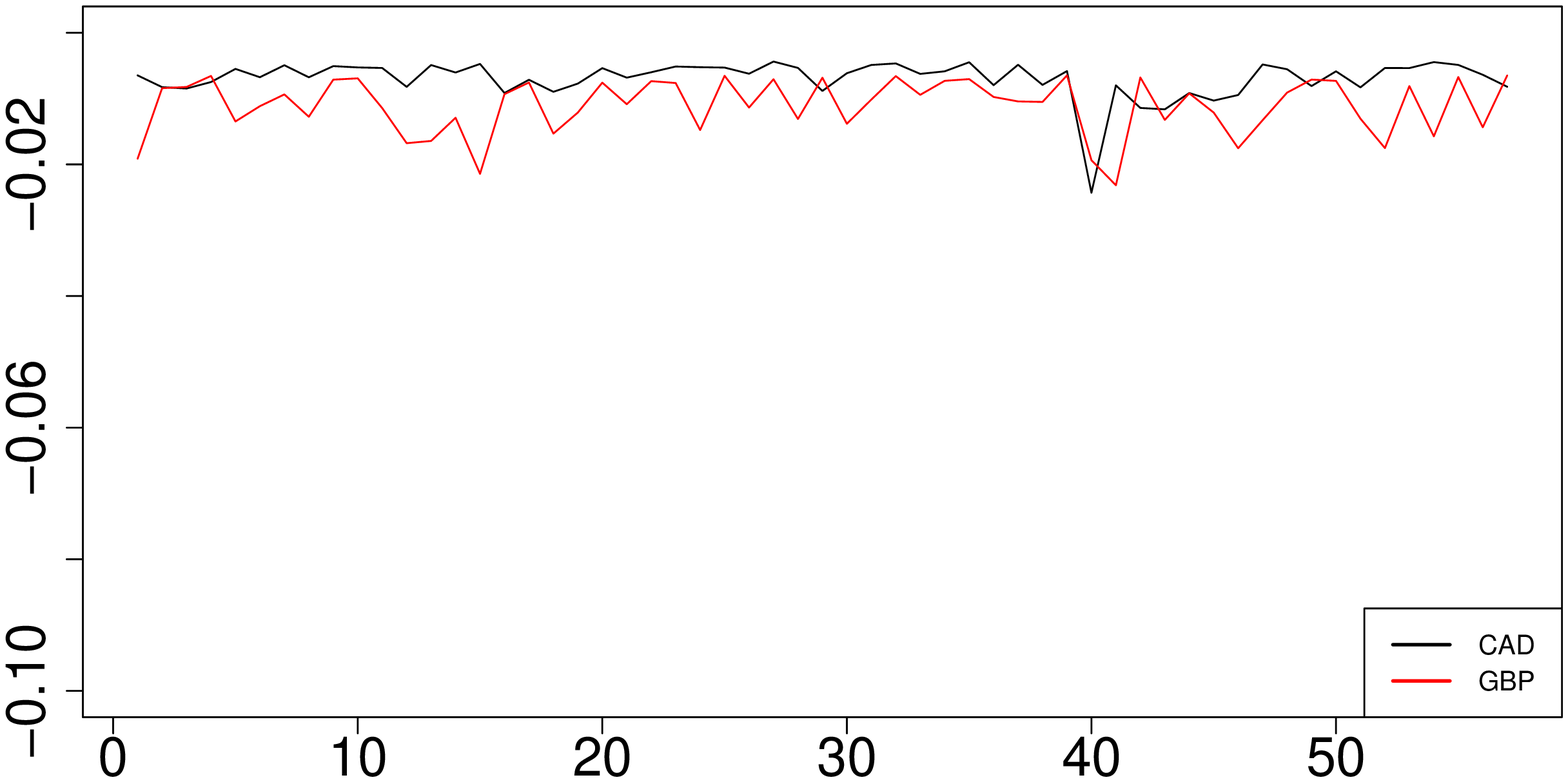}
        \caption{(CAD, GBP), $q = 0.1$, $m_{q} = 57$.}
        \label{Fig: CAD vs. GBP}
    \end{subfigure}
\caption{Original $x_t$'s (left-hand panels) and the pairs of extreme pseudo-observations (right-hand panels) for foreign currency exchange rates from January 4, 1971, to October 25, 2019.}
\label{fig:Log-returns}
\end{figure}

\begin{figure}[h!]
\centering
    \begin{subfigure}[b]{0.48\textwidth}
        \includegraphics[width=\textwidth]{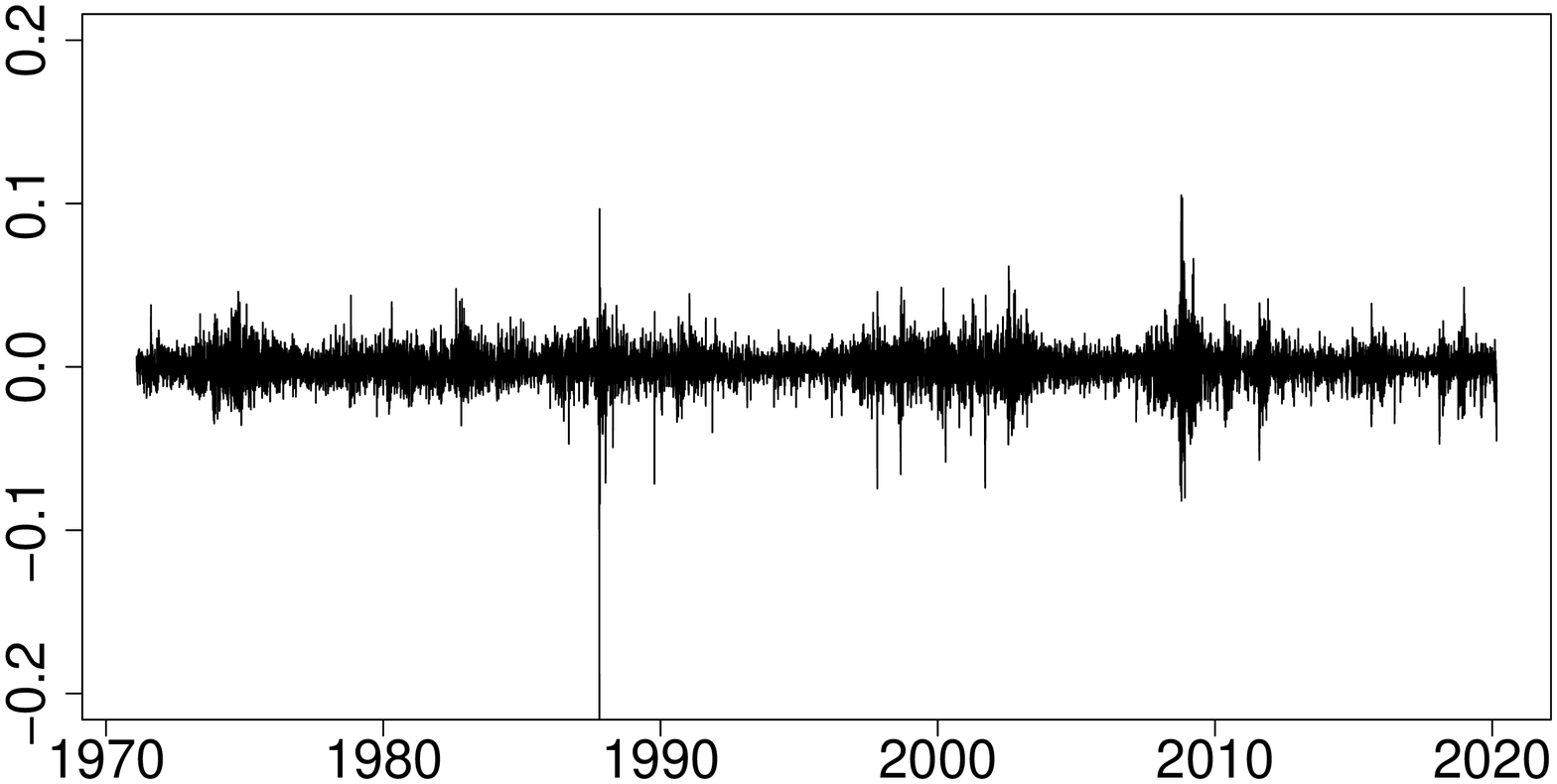}
        \caption{Dow Jones}
        \label{STOCK-DowJones}
    \end{subfigure}
\hfill
    \begin{subfigure}[b]{0.48\textwidth}
        \includegraphics[width=\textwidth]{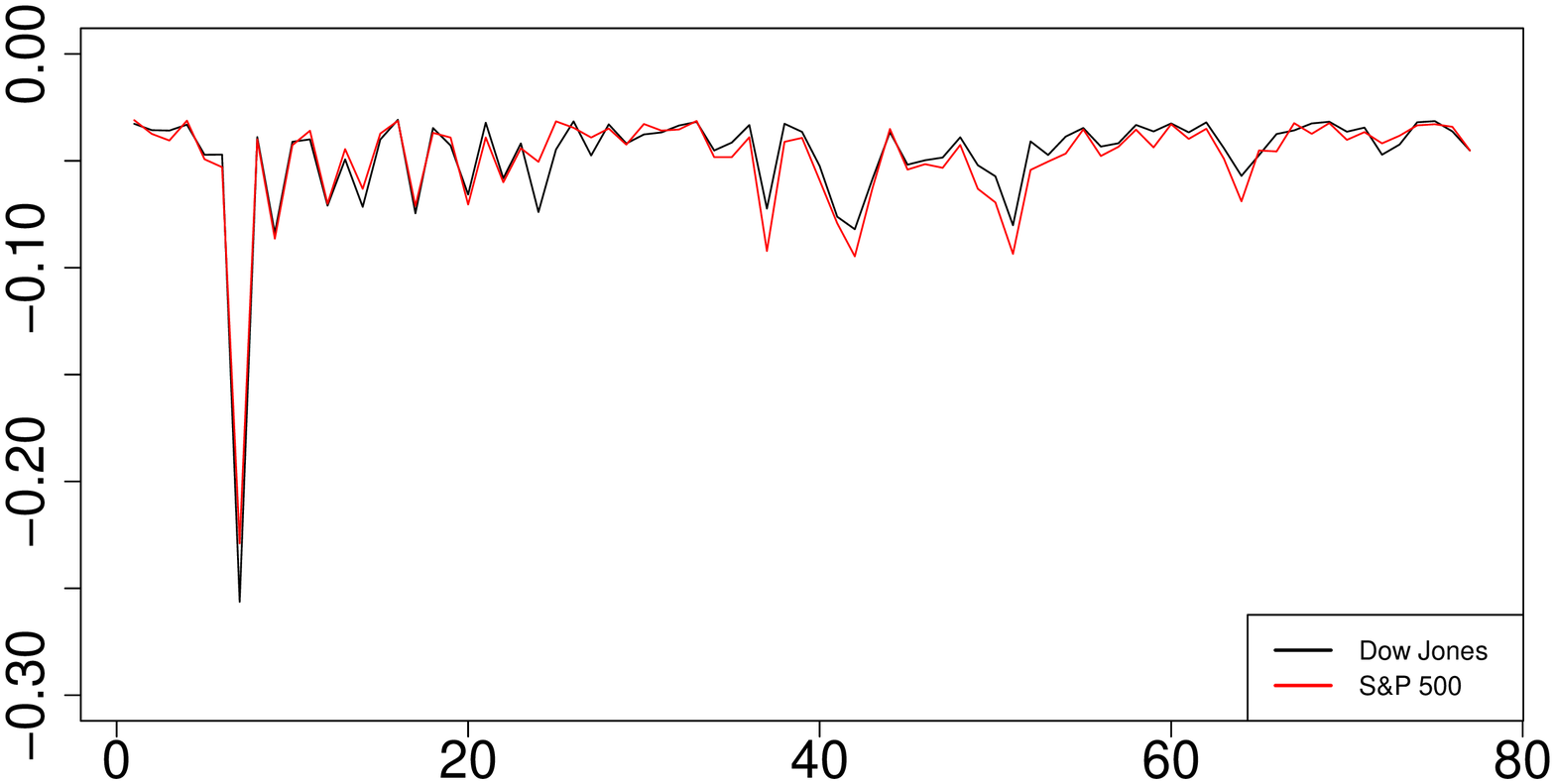}
        \caption{(Dow Jones, S\&P 500), $q = 0.0075$, $m_{q} = 77$.}
        \label{Fig: Dow Jones vs. SP500 00075}
    \end{subfigure}
\\
    \begin{subfigure}[b]{0.48\textwidth}
        \includegraphics[width=\textwidth]{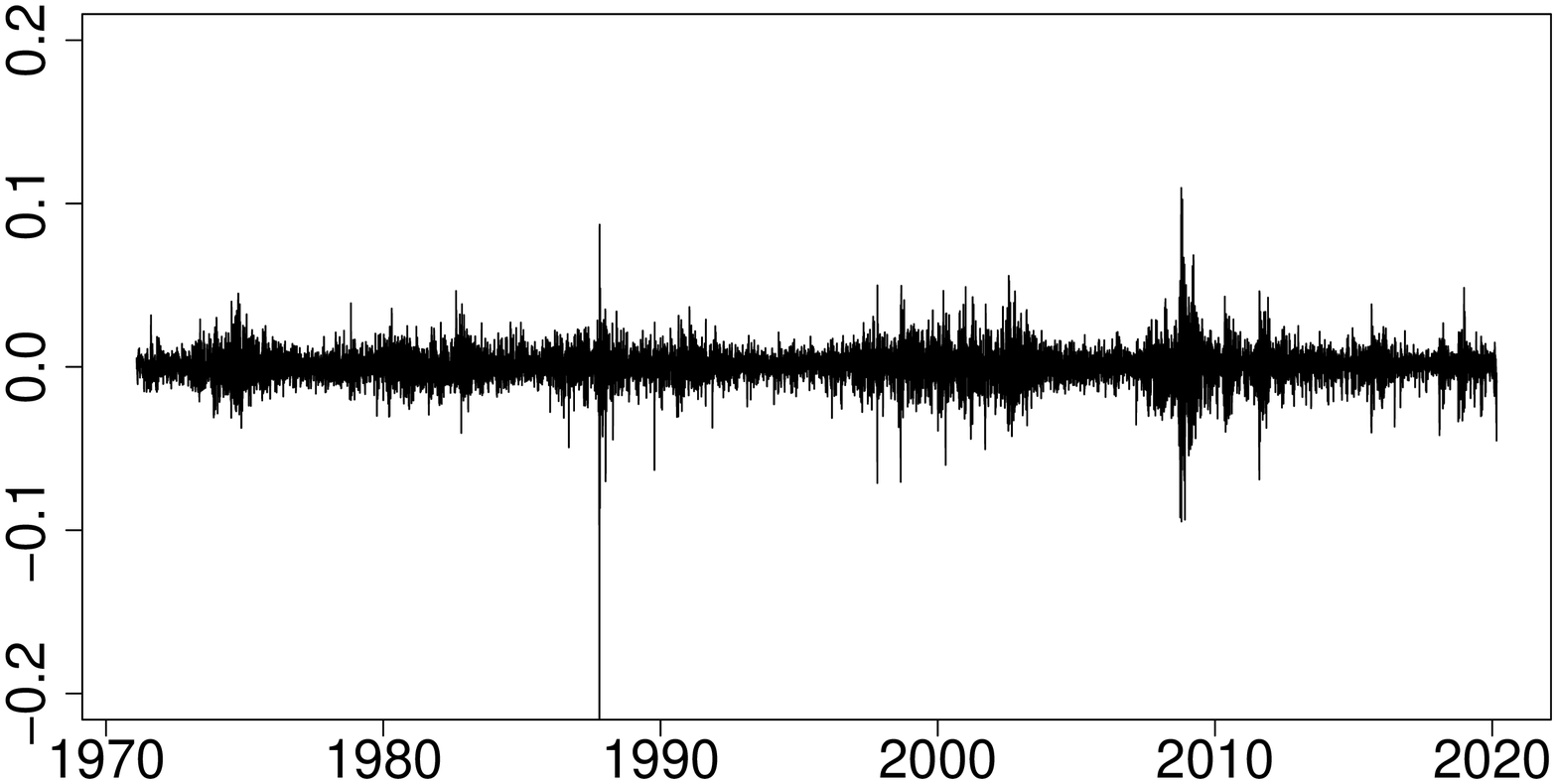}
        \caption{S\&P 500}
        \label{STOCK-SP500}
    \end{subfigure}
\hfill
    \begin{subfigure}[b]{0.48\textwidth}
        \includegraphics[width=\textwidth]{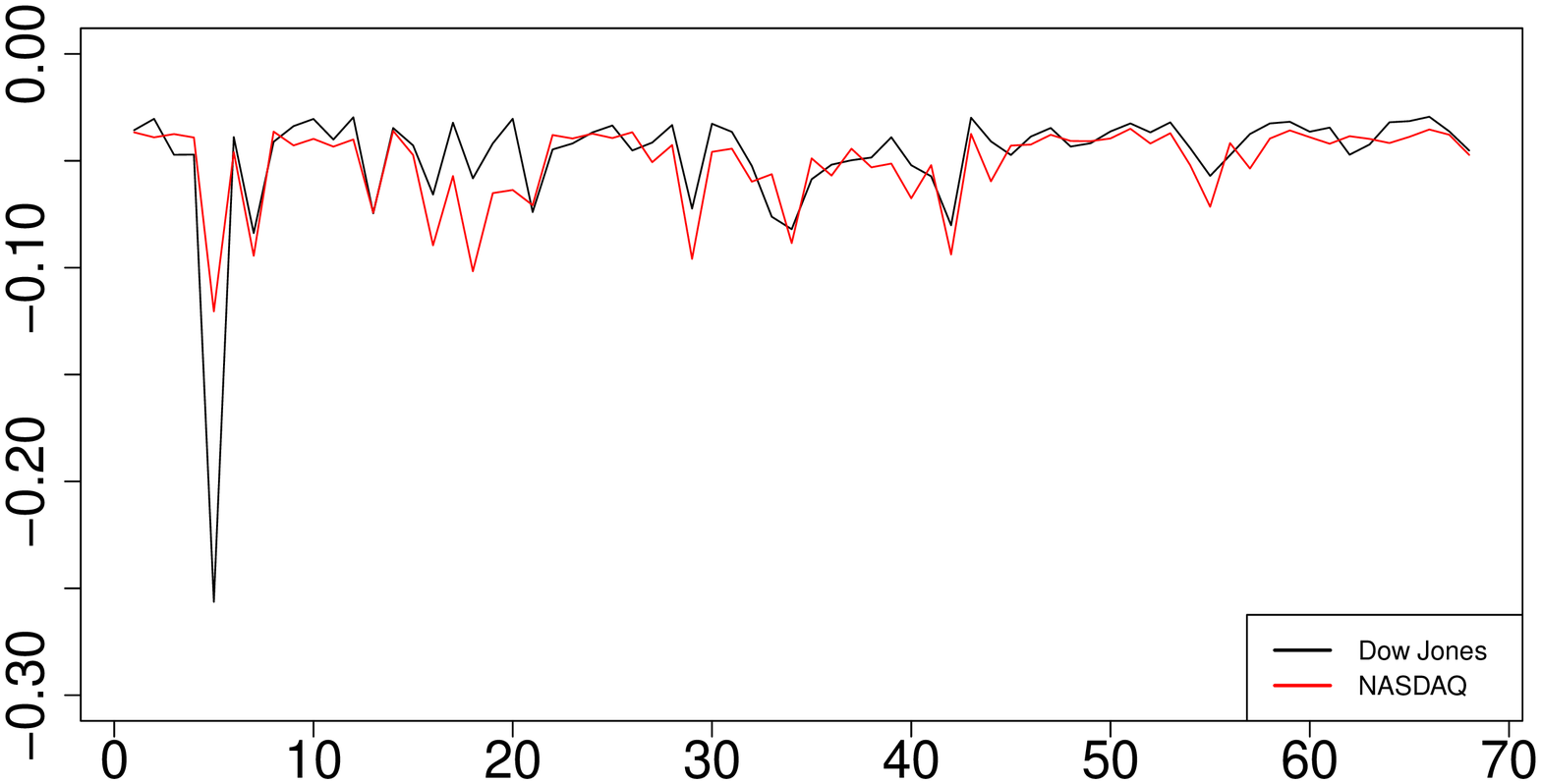}
        \caption{(Dow Jones, NASDAQ), $q = 0.01$, $m_{q} = 68$.}
        \label{Fig: Dow Jones vs. NASDAQ 001}
    \end{subfigure}
\\
    \begin{subfigure}[b]{0.48\textwidth}
        \includegraphics[width=\textwidth]{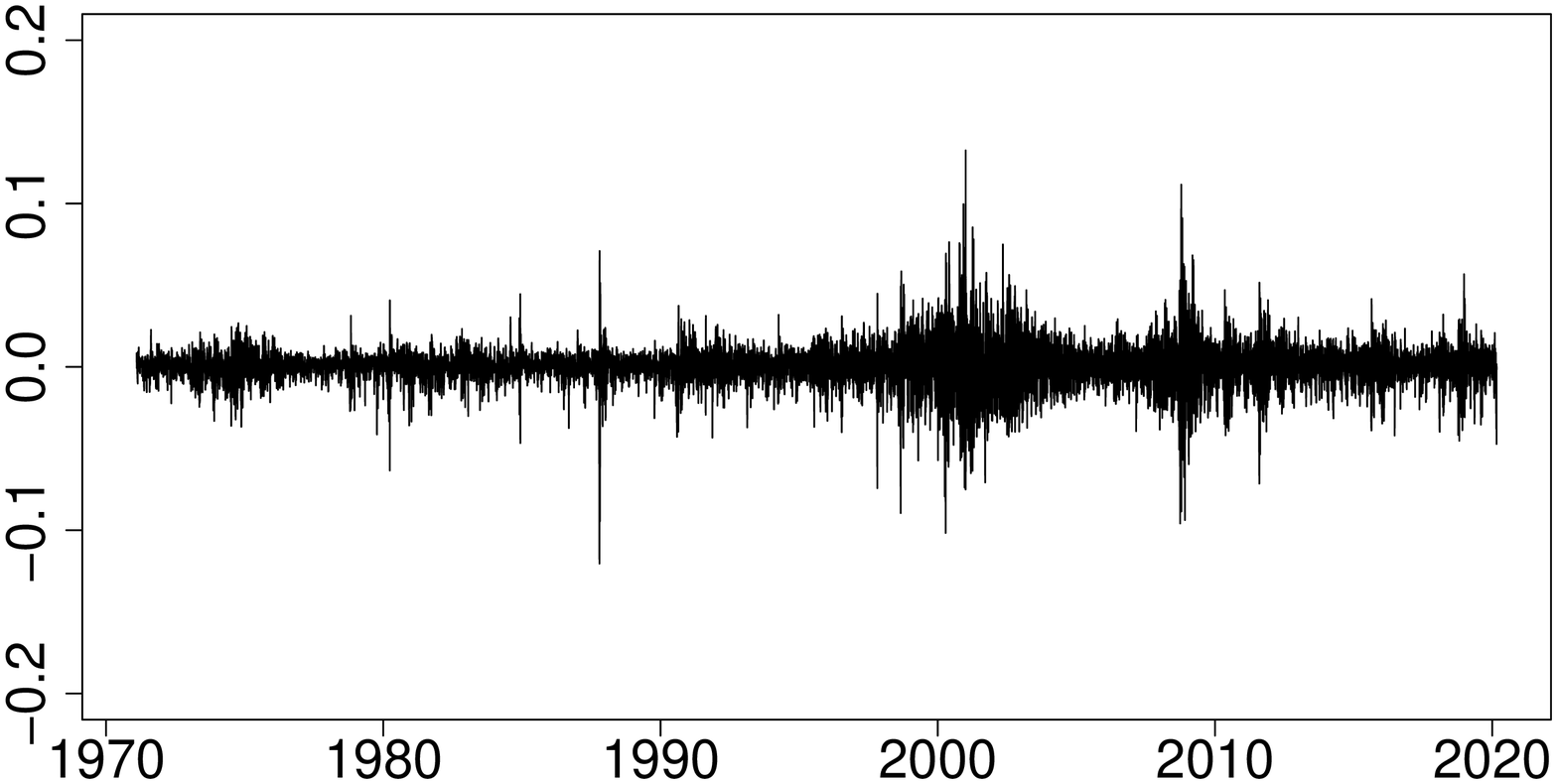}
        \caption{NASDAQ}
        \label{STOCK-NASDAQ}
    \end{subfigure}
\hfill
    \begin{subfigure}[b]{0.48\textwidth}
        \includegraphics[width=\textwidth]{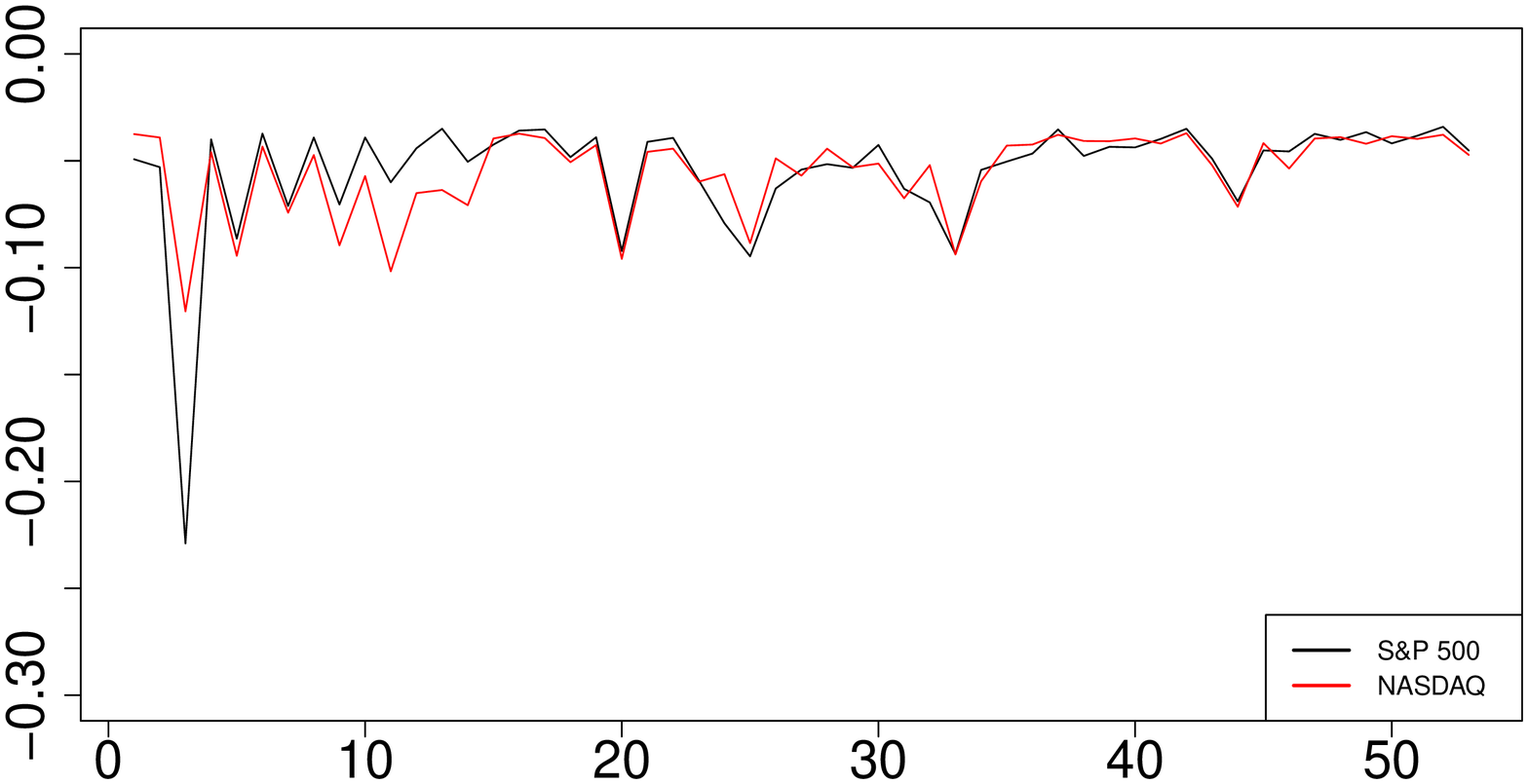}
        \caption{(S\&P 500, NASDAQ), $q = 0.0075$, $m_{q} = 53$.}
        \label{Fig: SP500 vs. NASDAQ 0075}
    \end{subfigure}
\caption{Original $x_t$'s (left-hand panels) and the pairs of extreme pseudo-observations (right-hand panels) for stock market indices from January 4, 1971, to February 28, 2020.}
\label{fig:STOCK-Log-returns}
\end{figure}

\begin{figure}[h!]
\centering
    \begin{subfigure}[b]{0.48\textwidth}
        \includegraphics[width=\textwidth]{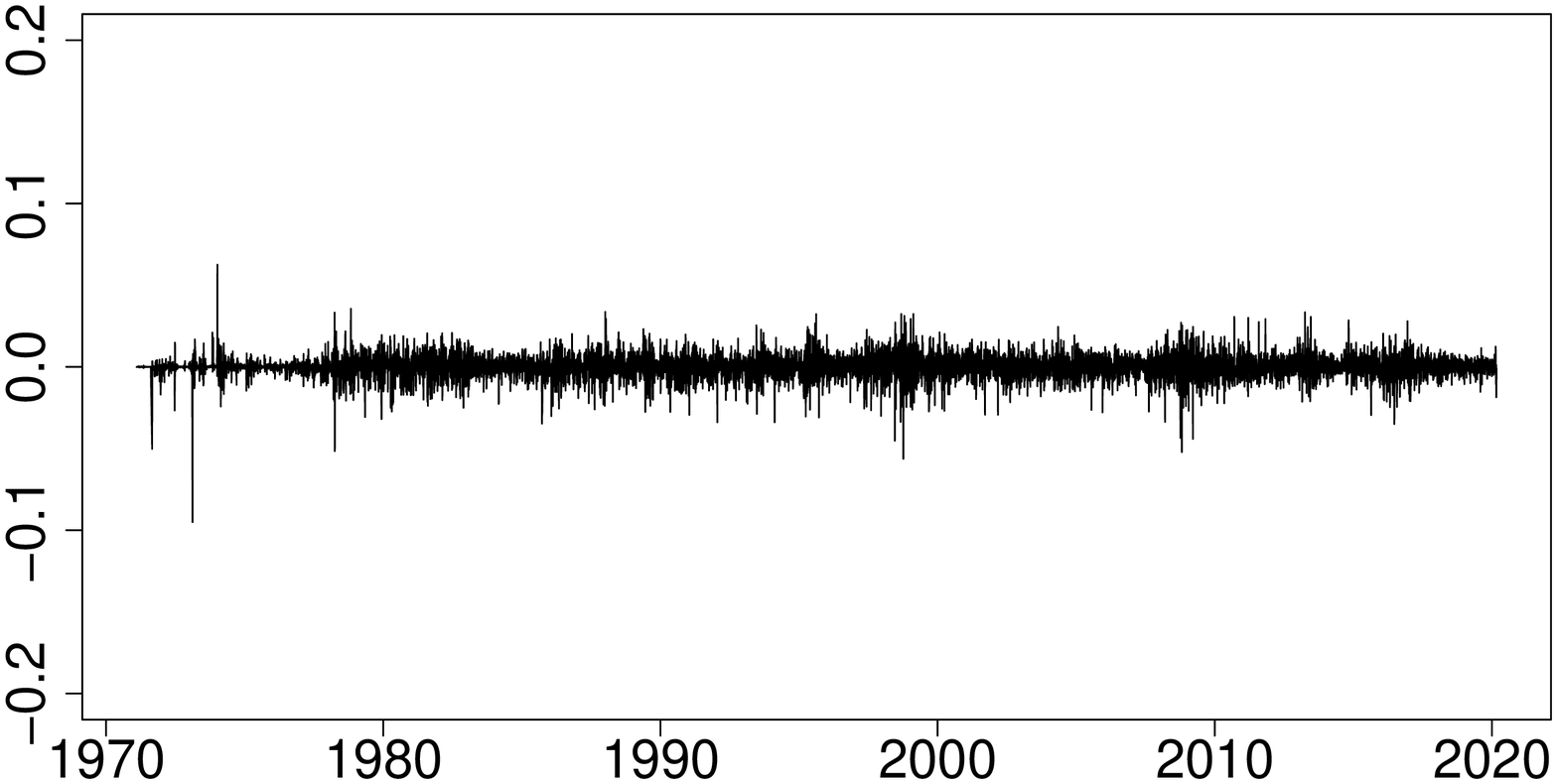}
        \caption{JPY/USD}
        \label{MIX-JPY}
    \end{subfigure}
\hfill
    \begin{subfigure}[b]{0.48\textwidth}
        \includegraphics[width=\textwidth]{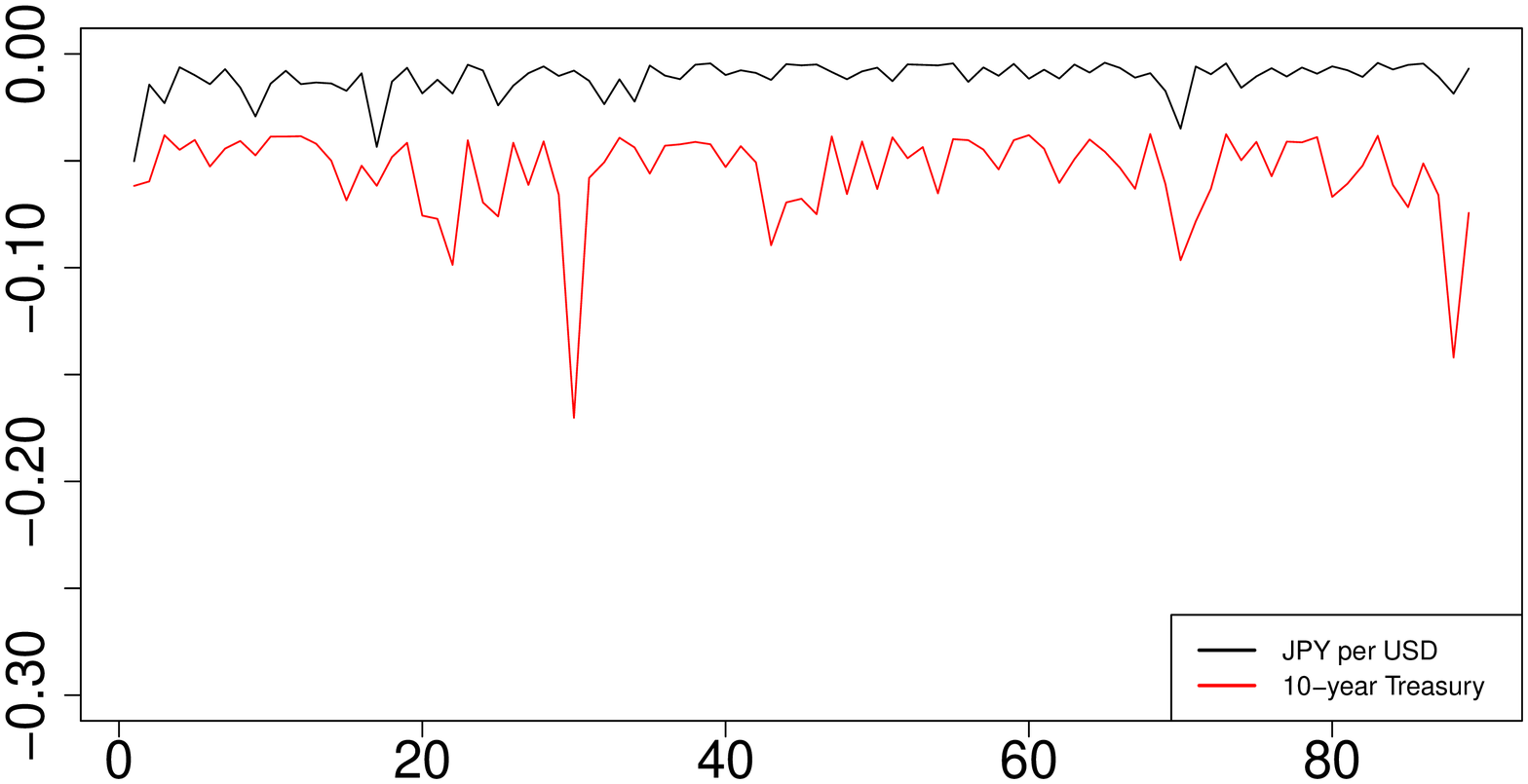}
        \caption{(JPY/USD, US10YT), $q = 0.05$, $m_{q} = 89$.}
        \label{Fig: JPY per USD vs. 10-year Treasury}
    \end{subfigure}
\\
    \begin{subfigure}[b]{0.48\textwidth}
        \includegraphics[width=\textwidth]{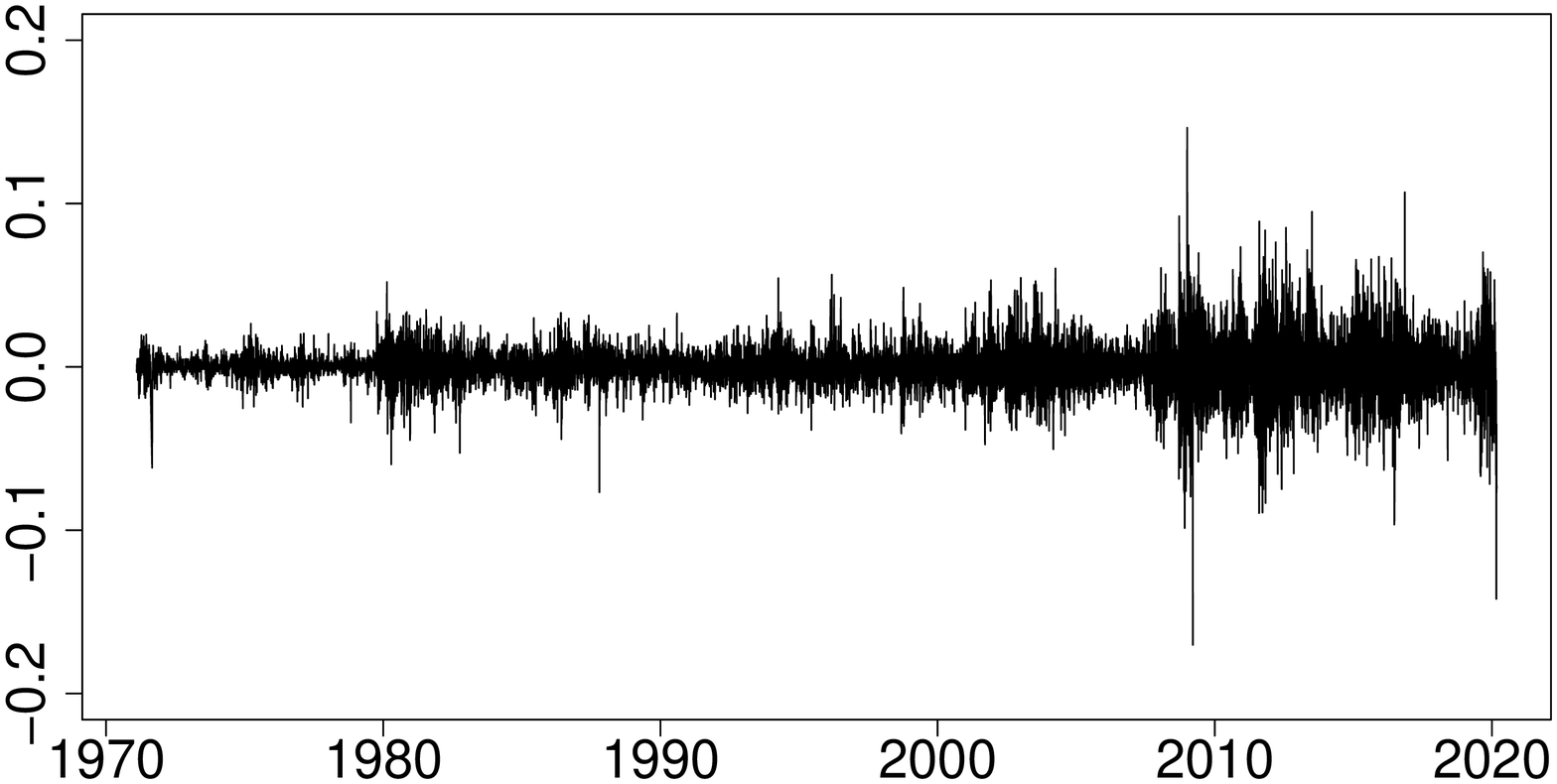}
        \caption{US10YT}
        \label{MIX-SP500}
    \end{subfigure}
\hfill
    \begin{subfigure}[b]{0.48\textwidth}
        \includegraphics[width=\textwidth]{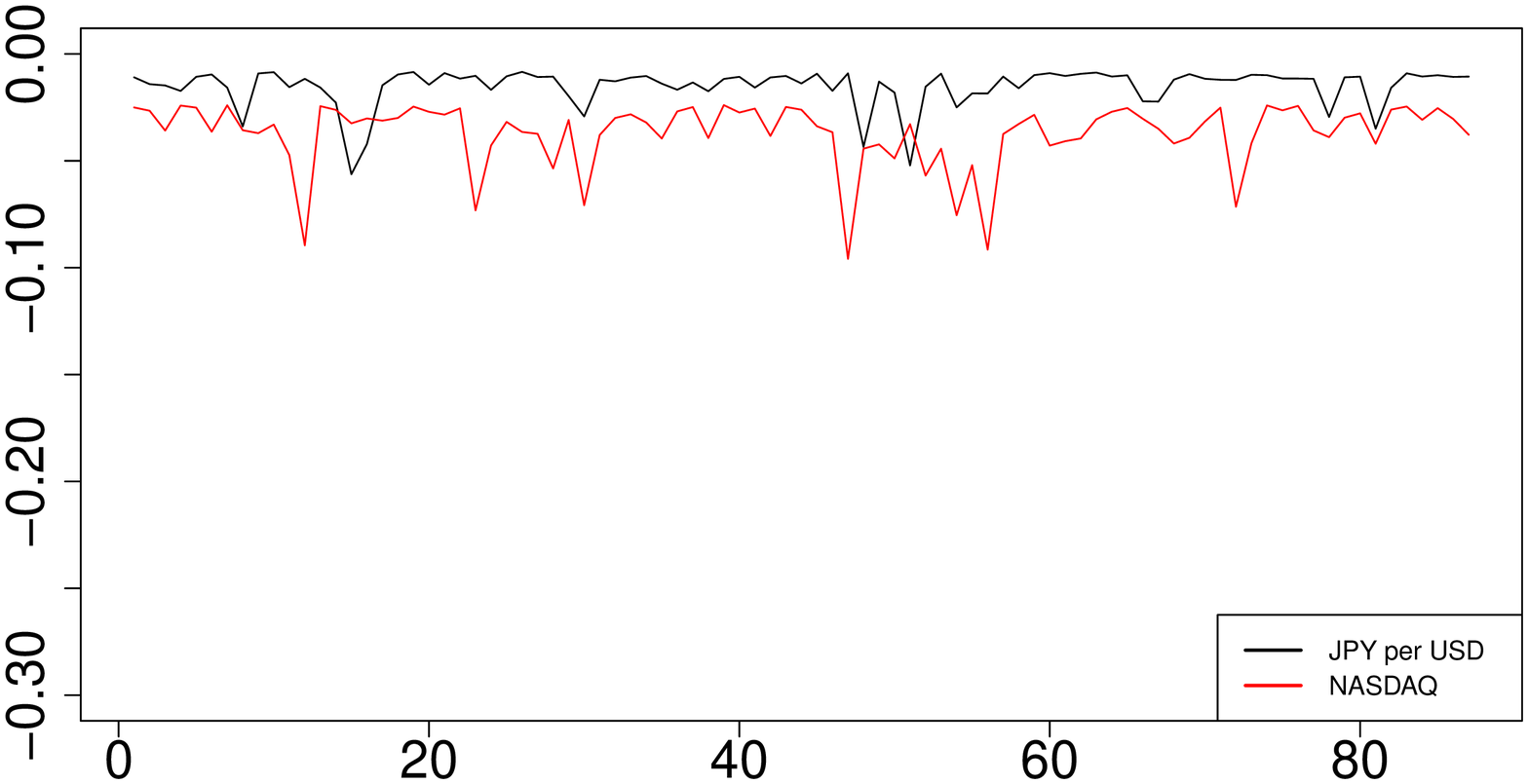}
        \caption{(JPY/USD, NASDAQ), $q = 0.05$, $m_{q} = 87$.}
        \label{Fig: JPY per USD vs. NASDAQ}
    \end{subfigure}
\\
    \begin{subfigure}[b]{0.48\textwidth}
        \includegraphics[width=\textwidth]{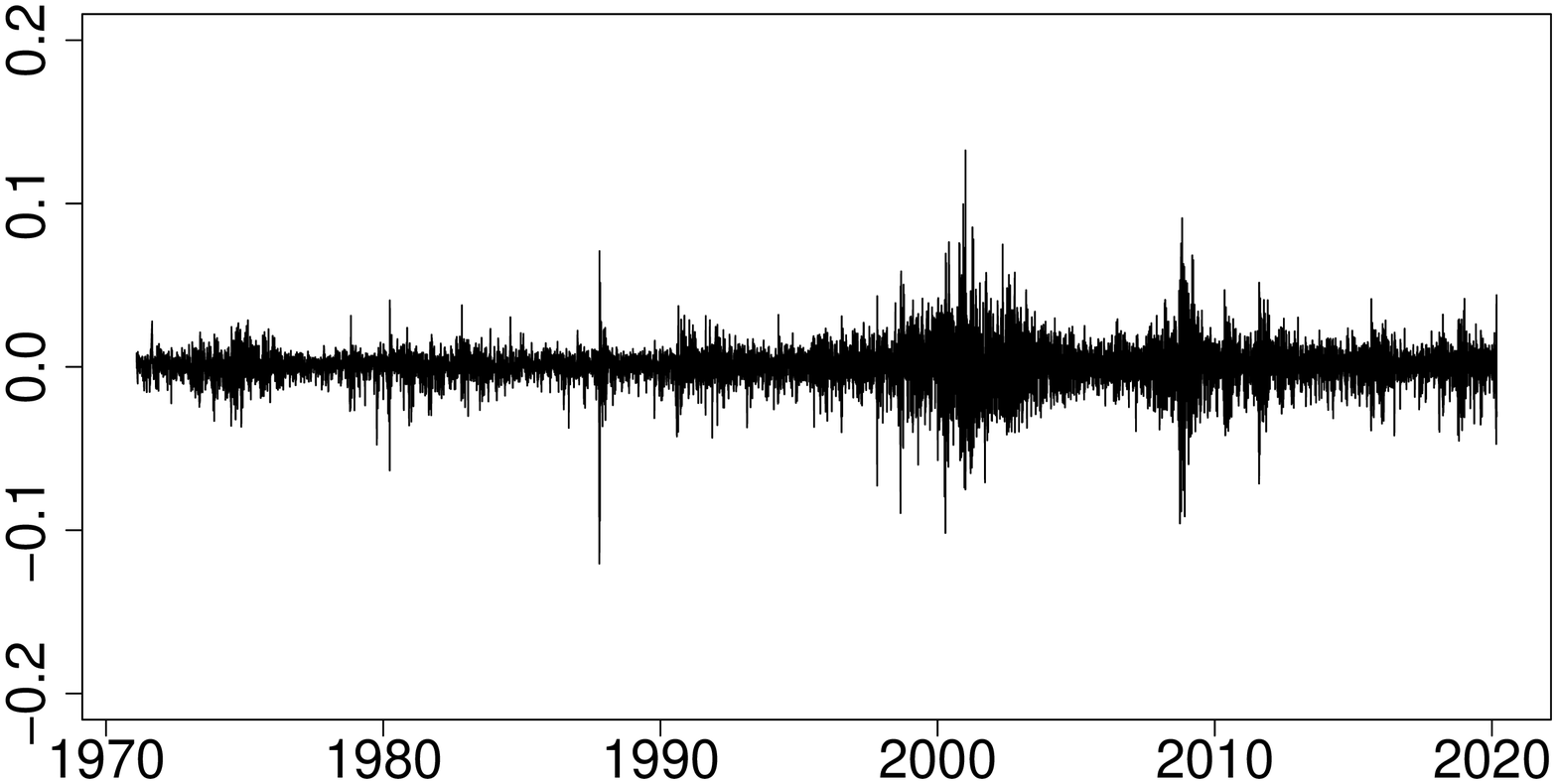}
        \caption{NASDAQ}
        \label{MIX-NASDAQ}
    \end{subfigure}
\hfill
    \begin{subfigure}[b]{0.48\textwidth}
        \includegraphics[width=\textwidth]{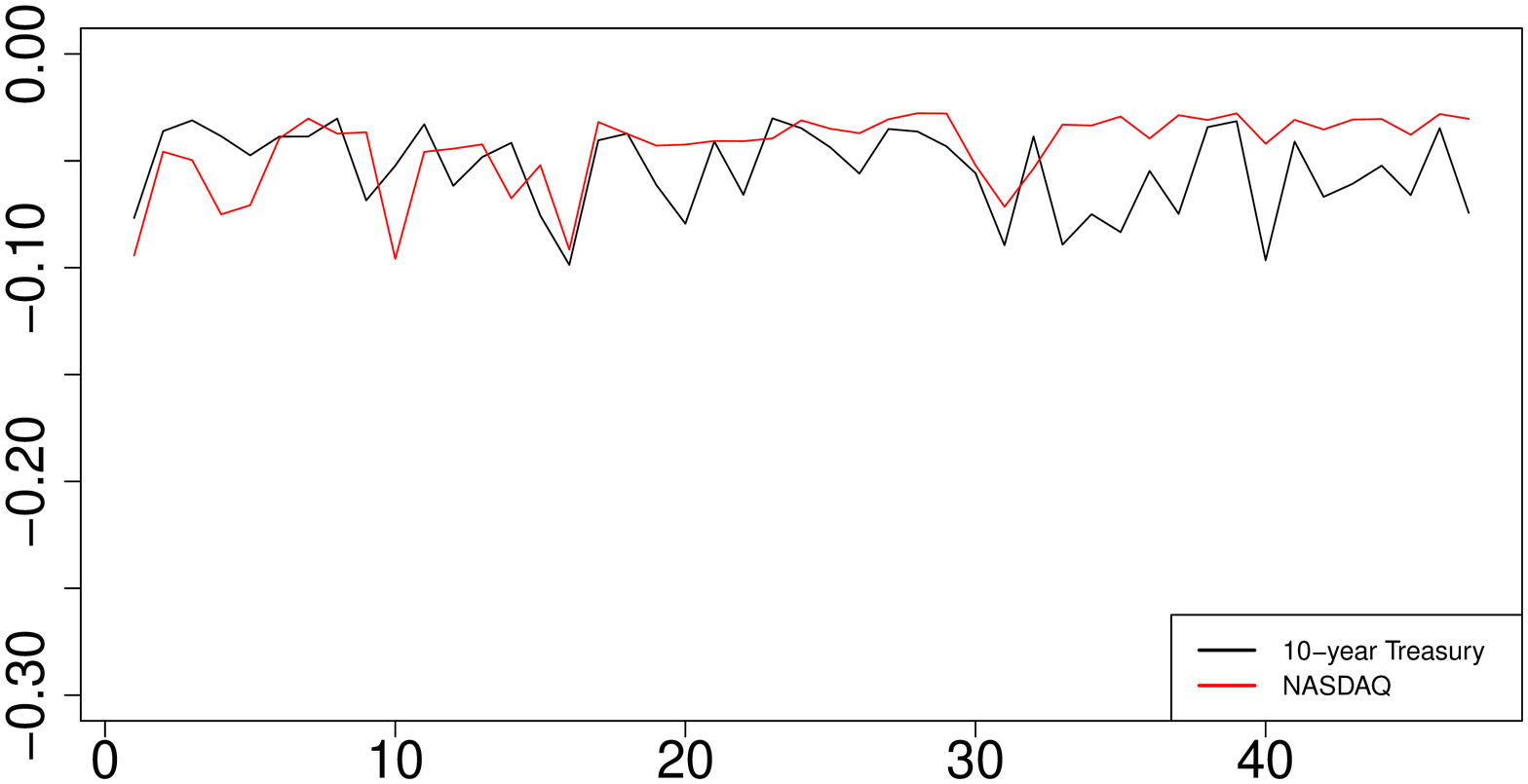}
        \caption{(US10YT, NASDAQ), $q = 0.025$, $m_{q} = 47$.}
        \label{Fig: 10-year Treasury vs. NASDAQ}
    \end{subfigure}
\caption{Original $x_t$'s (left-hand panels) and the pairs of extreme pseudo-observations (right-hand panels) for diverse financial instruments from February 5, 1971, to March 3, 2020.}
\label{MIX-fig:Log-returns}
\end{figure}

In Figures~\ref{fig:Log-returns}--\ref{MIX-fig:Log-returns} we depict the differenced log-time-series $x_t=\log(x^0_t)-\log(x^0_{t-1})$ (left-hand panels) and the extreme pseudo-observations (right-hand panels) that arise from the time series data specified in Section~\ref{appl-0}.
With the thresholds $q\in (0,1)$ reported in Table~\ref{pvalues}, %
\begin{table}[h!]
\centering
\begin{tabular}{ lccccccc }
\hline\hline
  & \multicolumn{7}{c}{Threshold $q$}
  \\
\cmidrule(r{5pt}){2-8}
 Pairs                & 0.0075 & 0.01 & 0.025 & 0.05 & 0.075 & 0.085 & 0.1  \\
\hline
(JPY, CAD)            & ... & ... & ... & ... & 100(123) & ... & ...         \\
(JPY, GBP)            & ... & ... & ... & ... & ... & 90(64)       & ...         \\
(CAD, GBP)            & ... & ... & ... & ... & ... & ... & 95(57) \\
\hline
(Dow Jones, S\&P 500) & 60(77)   & ... & ... & ... & ... & ... & ...   \\
(Dow Jones, NASDAQ)   & ... & 77(68) & ... & ... & ... & ... & ...   \\
(S\&P 500, NASDAQ)    &  76(53)  & ... & ... & ... & ... & ... & ...   \\
  \hline
(JPY/USD, US10YT)     & ... & ... & ... & 100(89) & ... & ... & ...   \\
(JPY/USD, NASDAQ)     & ... & ... & ... & 100(87) & ... & ... & ...   \\
(US10YT, NASDAQ)      & ... & ... & 88(47) & ... & ... & ... & ...   \\
  \hline
\end{tabular}
\caption{The percentages of $p$-values retaining the null of white noise at the significance level $\alpha=0.05$ alongside the sample sizes $m_{q}$ (in parentheses) for appropriate choices of $q$.}
\label{pvalues}
\end{table}
the time series give rise to paired extreme pseudo-observations  that resemble a white noise; see the right-hand panels of Figures~\ref{fig:Log-returns}--\ref{MIX-fig:Log-returns}. To substantiate this claim, we run several portmanteau tests for the null hypothesis
$$
\mathcal{H}_0: \boldsymbol{\Gamma}_{L} = \mathbf{0},\quad L = 1,\ldots,20,
$$
where $\boldsymbol{\Gamma}_{L} = \mathrm{Cov}(\boldsymbol{\varepsilon}_t,\boldsymbol{\varepsilon}_{t-L})$ and  $\boldsymbol{\varepsilon}_t$, $t=1,\dots , m_q$, are the residuals obtained by fitting the original data to the time series model VARMA for sufficiently many lags \citep{Mahdi2012}. The selected portmanteau tests include those of \citet{Mahdi2012}, \citet{Box1970}, \citet{Ljung1978}, \citet{Hosking1980}, and \citet{Li1981}. The percentages of $p$-values above the 5\% significance level (meaning that the null of white noise is retained) are given in Table~\ref{pvalues}, where we also report the sample sizes $m_{q}$.

The different choices of $q\in (0,1)$ warrant an explanation. First, we want to work with as small $q>0$ as possible, mainly due to two reasons:
\begin{itemize}
\item
the estimator's deterministic bias becomes small (recall Theorem~\ref{thm: essinf}),
\item the time series of extreme pseudo-observations becomes nearly a white-noise.
\end{itemize}
Working close to a white noise is useful as it helps to reliably calculate critical values of the hypothesis tests for bound~\eqref{pqd}, which we need for the use of Theorem~\ref{thm: essinf}.

\subsection{Testing the validity of bound~\eqref{pqd}}
\label{append-3a}

For the real time series that we are exploring, we want to statistically test the reasonableness of bound~\eqref{pqd}. For this, we adapt the Kolmogorov-Smirnov (K-S), Cram\'{e}r-von Mises (C-vM), Anderson-Darling (A-D) one-sided statistics \citep{Tang2019}:
\begin{gather}
\sqrt{m_{q}}\sup_{(u,v)\in[0,1]\times [0,1]}\{uv-F^*_{q,\mathcal{M}_q}(u,v)\}_{+} ,
\label{ks-pqd}
\\
m_{q}\int_{[0,1]\times [0,1]}\big(\{uv-F^*_{q,\mathcal{M}_q}(u,v)\}_{+}\big)^2\dif F^*_{q,\mathcal{M}_q}(u,v) ,
\label{cm-pqd}
\\
m_{q}\int_{[0,1]\times [0,1]}\dfrac{\big(\{uv-F^*_{q,\mathcal{M}_q}(u,v)\}_{+}\big)^2}{u(1-u)v(1-v)}\dif F^*_{q,\mathcal{M}_q}(u,v),
\label{ad-pqd}
\end{gather}
respectively, where $F^*_{q,\mathcal{M}_q}$ is defined by equation~\eqref{def-fqm}.
Specifically, we use these three statistics to test the null $H_0$ of having bound~\eqref{pqd} versus the alternative $H_1$ of not having the bound:
\begin{align*}
H_0: &\quad  \text{$F_q^*(u,v)\ge uv$ for all $(u,v)\in [0,1]\times [0,1]$}
\\
H_1: &\quad  \text{$F_q^*(u,v)< uv$ for some $(u,v)\in [0,1]\times [0,1]$}
\end{align*}

\begin{note}
The null $H_0$ can be reformulated as $F_q^*\ge C^{\perp}$ and the alternative $H_1$ as $F_q^*\ngeq C^{\perp}$.
\end{note}

The critical values of the tests are obtained by sampling from the pairs of pseudo observations. Namely, we calculate the test statistics,
repeat the procedure $N = 10,000$ times, obtain so many values of the test statistics, and finally calculate the 95$^{\rm th}$ percentiles of the respective test-statistic values. The decision rule is to retain the null $H_0$ if the test statistic is smaller than the critical value, and to reject it otherwise. The obtained results are summarized in Tables~\ref{Test: PQD-appl-1}--\ref{Test: PQD-appl-3}, where the abbreviations ``Stat,'' ``Crit,'' and ``Deci'' stand for the test statistic value, the critical value, and the decision, respectively.%
\begin{table}[h!]
\centering
\begin{tabular}{ lccccccccc }
\hline\hline
  & \multicolumn{3}{c}{(JPY, CAD)} & \multicolumn{3}{c}{(JPY, GBP)} & \multicolumn{3}{c}{(CAD, GBP)}
  \\
  & \multicolumn{3}{c}{$q=0.075$, $m_{q}=123$} & \multicolumn{3}{c}{$q=0.085$, $m_{q}=64$} & \multicolumn{3}{c}{$q=0.1$, $m_{q}=57$}
  \\
  \cmidrule(r{5pt}){2-4} \cmidrule(r{5pt}){5-7} \cmidrule(r{5pt}){8-10}
  Test & Stat & Crit & Deci & Stat & Crit & Deci & Stat & Crit & Deci
  \\
  \hline
 K-S  & 0.0189 & 1.2119 & $H_0$ & 0.0404 & 1.1195 & $H_0$ & 0.5781 & 1.0989 & $H_0$ \\
 C-vM & 0.0000 & 0.2013 & $H_0$ & 0.0000 & 0.1844 & $H_0$ & 0.0330 & 0.1737 & $H_0$ \\
 A-D  & 0.0008 & 28.7139 & $H_0$ & 0.0064 & 23.9732 & $H_0$ & 6.0110 & 24.5422 & $H_0$ \\
  \hline
\end{tabular}
\caption{Testing $H_0$ vs $H_1$ of pseudo observations of foreign currency exchange rates.}
\label{Test: PQD-appl-1}
\end{table}
\begin{table}[h!]
\centering
\begin{tabular}{ lrccrccrcc }
\hline\hline
   & \multicolumn{3}{c}{(Dow Jones, S\&P 500)} & \multicolumn{3}{c}{(Dow Jones, NASDAQ)} & \multicolumn{3}{c}{(S\&P 500, NASDAQ)}
  \\
   & \multicolumn{3}{c}{$q=0.0075$, $m_{q}=77$} & \multicolumn{3}{c}{$q=0.01$, $m_{q}=68$} & \multicolumn{3}{c}{$q=0.0075$, $m_{q}=53$}
  \\
  \cmidrule(r{5pt}){2-4} \cmidrule(r{5pt}){5-7} \cmidrule(r{5pt}){8-10}
  Test & Stat & Crit & Deci & Stat & Crit & Deci & Stat & Crit & Deci
  \\
  \hline
 K-S  & 0.0000 & 1.1474 & $H_0$ & 0.0000 & 1.1350 & $H_0$ & 0.0000 & 1.0905 & $H_0$ \\
 C-vM & 0.0000 & 0.1922 & $H_0$ & 0.0000 & 0.1900 & $H_0$ & 0.0000 & 0.1786 & $H_0$ \\
 A-D  & 0.0000 & 26.5523 & $H_0$ & 0.0000 & 24.7340 & $H_0$ & 0.0000 & 22.9812 & $H_0$ \\
  \hline
\end{tabular}
\caption{Testing $H_0$ vs $H_1$ of pseudo observations of stock market indices.}
\label{Test: PQD-appl-2}
\end{table}
\begin{table}[h!]
\centering
\begin{tabular}{ lccccccccc }
\hline\hline
   & \multicolumn{3}{c}{(JPY/USD, US10YT)} & \multicolumn{3}{c}{(JPY/USD, NASDAQ)} & \multicolumn{3}{c}{(US10YT, NASDAQ)}
  \\
   & \multicolumn{3}{c}{$q=0.05$, $m_{q}=89$} & \multicolumn{3}{c}{$q=0.05$, $m_{q}=87$} & \multicolumn{3}{c}{$q=0.025$, $m_{q}=47$}
  \\
  \cmidrule(r{5pt}){2-4} \cmidrule(r{5pt}){5-7} \cmidrule(r{5pt}){8-10}
  Test & Stat & Crit & Deci & Stat & Crit & Deci & Stat & Crit & Deci
  \\
  \hline
 K-S  & 0.0000 & 1.1715 & $H_0$ & 0.1913 & 1.1766 & $H_0$ & 0.0000 & 1.0739 & $H_0$ \\
 C-vM & 0.0000 & 0.1889 & $H_0$ & 0.0007 & 0.2004 & $H_0$ & 0.0000 & 0.1794 & $H_0$ \\
 A-D  & 0.0000 & 25.2966 & $H_0$ & 0.6637 & 26.5172 & $H_0$ & 0.0000 & 23.8698 & $H_0$ \\
  \hline
\end{tabular}
\caption{Testing $H_0$ vs $H_1$ of pseudo observations of diverse financial instruments.}
\label{Test: PQD-appl-3}
\end{table}
The decision is to retain the null $H_0$ when the statistic is smaller than the critical value. In every case, the three tests retain the null $H_0$.

To gain an additional insight, in Section~\ref{append-3} we test the null of the equation $F_q^*(u,v)= uv$ for all $(u,v)\in [0,1]\times [0,1]$, which is the  ``boundary'' of the null $H_0$ introduced earlier.

\subsection{Testing the boundary case of bound~\eqref{pqd}}
\label{append-3}

In the examples of Section~\ref{append-3a}, all of which retained the null $H_0$, we now statistically test the reasonableness of the boundary case  $F_q^*(u,v)= uv$ for all $(u,v)\in [0,1]\times [0,1]$. We adapt the Kolmogorov-Smirnov (K-S), Cram\'{e}r-von Mises (C-vM), Anderson-Darling (A-D) one-sided statistics \citep[cf.][]{Tang2019}
\begin{gather}
\sqrt{m_{q}}\sup_{(u,v)\in[0,1]\times [0,1]}\{F^*_{q,\mathcal{M}_q}(u,v) - uv\}_{+} ,
\label{ks-spqd}
\\
m_{q}\int_{[0,1]\times [0,1]}\big(\{F^*_{q,\mathcal{M}_q}(u,v) - uv\}_{+}\big)^2\dif F^*_{q,\mathcal{M}_q}(u,v),
\label{cm-spqd}
\\
m_{q}\int_{[0,1]\times [0,1]}\dfrac{\big(\{F^*_{q,\mathcal{M}_q}(u,v) - uv\}_{+}\big)^2}{u(1-u)v(1-v)}\dif F^*_{q,\mathcal{M}_q}(u,v),
\label{ad-spqd}
\end{gather}
respectively. Specifically, we use them to test the hypotheses
\begin{align*}
H_0^*: &\quad  \text{$F_q^*(u,v)= uv$ for all $(u,v)\in [0,1]\times [0,1]$}
\\
H_1^*: &\quad  \text{$F_q^*(u,v)> uv$ for some $(u,v)\in [0,1]\times [0,1]$}
\end{align*}

\begin{note}
The null $H_0^*$ can be viewed as the ``boundary'' $F_q^*=C^{\perp}$ of the null $H_0$ introduced in Appendix~\ref{append-3a}.
\end{note}

Coming now back to our main discussion, we note that the procedures for calculating the critical values for any of the three tests for $H_0^*$ vs $H_1^*$ using statistics \eqref{ks-spqd}--\eqref{ad-spqd} are analogous to those we used in Section~\ref{append-3a} for $H_0$ vs $H_1$. The decision to retain the null $H_0^*$ is, of course, taken when the statistic is smaller than the critical value. Our findings are summarized in Tables~\ref{Test: SPQD-appl-1}--\ref{Test: SPQD-appl-2}. %
\begin{table}[h!]
\centering
\begin{tabular}{ lccccccccc }
\hline\hline
  & \multicolumn{3}{c}{(JPY, CAD)} & \multicolumn{3}{c}{(JPY, GBP)} & \multicolumn{3}{c}{(CAD, GBP)}
  \\
  & \multicolumn{3}{c}{$q=0.075$, $m_{q}=123$} & \multicolumn{3}{c}{$q=0.085$, $m_{q}=64$} & \multicolumn{3}{c}{$q=0.1$, $m_{q}=57$}
  \\
  \cmidrule(r{5pt}){2-4} \cmidrule(r{5pt}){5-7} \cmidrule(r{5pt}){8-10}
  Test & Stat & Crit & Deci & Stat & Crit & Deci & Stat & Crit & Deci
  \\
  \hline
 K-S  & 1.4859 & 1.3546 & $H_1^*$ & 1.4837 & 1.3386 & $H_1^*$ & 0.9139 & 1.3231 & $H_0^*$ \\
 C-vM & 0.5874 & 0.2969 & $H_1^*$ & 0.6968 & 0.3198 & $H_1^*$ & 0.0782 & 0.3299 & $H_0^*$ \\
 A-D  & 25.6607 & 44.9833 & $H_0^*$ & 91.2540 & 50.1569 & $H_1^*$ & 4.0293 & 49.6459 & $H_0^*$ \\
  \hline
\end{tabular}
\caption{Testing $H_0^*$ vs $H_1^*$ of pseudo observations of foreign currency exchange rates.}
\label{Test: SPQD-appl-1}
\end{table}
\begin{table}[h!]
\centering
\begin{tabular}{ lrccrccrcc }
\hline\hline
   & \multicolumn{3}{c}{(Dow Jones, S\&P 500)} & \multicolumn{3}{c}{(Dow Jones, NASDAQ)} & \multicolumn{3}{c}{(S\&P 500, NASDAQ)}
  \\
   & \multicolumn{3}{c}{$q=0.0075$, $m_{q}=77$} & \multicolumn{3}{c}{$q=0.01$, $m_{q}=68$} & \multicolumn{3}{c}{$q=0.0075$, $m_{q}=53$}
  \\
  \cmidrule(r{5pt}){2-4} \cmidrule(r{5pt}){5-7} \cmidrule(r{5pt}){8-10}
  Test & Stat & Crit & Deci & Stat & Crit & Deci & Stat & Crit & Deci
  \\
  \hline
 K-S  & 2.4058 & 1.3441 & $H_1^*$ & 1.7760 & 1.3335 & $H_1^*$ & 1.6858 & 1.3208 & $H_1^*$ \\
 C-vM & 2.9460 & 0.3107 & $H_1^*$ & 1.3744 & 0.3269 & $H_1^*$ & 1.3100 & 0.3343 & $H_1^*$ \\
 A-D  & 90.9184 & 47.4610 & $H_1^*$ & 61.2180 & 49.3820 & $H_1^*$ & 51.2729 & 51.2274 & $H_1^*$ \\
  \hline
\end{tabular}
\caption{Testing $H_0^*$ vs $H_1^*$ of pseudo observations of stock market indices.}
\label{Test: SPQD-appl-2}
\end{table}
\begin{table}[h!]
\centering
\begin{tabular}{ lccccccccc }
\hline\hline
   & \multicolumn{3}{c}{(JPY/USD, US10YT)} & \multicolumn{3}{c}{(JPY/USD, NASDAQ)} & \multicolumn{3}{c}{(US10YT, NASDAQ)}
  \\
   & \multicolumn{3}{c}{$q=0.05$, $m_{q}=89$} & \multicolumn{3}{c}{$q=0.05$, $m_{q}=87$} & \multicolumn{3}{c}{$q=0.025$, $m_{q}=47$}
  \\
  \cmidrule(r{5pt}){2-4} \cmidrule(r{5pt}){5-7} \cmidrule(r{5pt}){8-10}
  Test & Stat & Crit & Deci & Stat & Crit & Deci & Stat & Crit & Deci
  \\
  \hline
 K-S  & 1.8927 & 1.3442 & $H_1^*$ & 0.8564 & 1.3465 & $H_0^*$ & 2.1463 & 1.3158 & $H_1^*$ \\
 C-vM & 1.0709 & 0.3092 & $H_1^*$ & 0.2266 & 0.3142 & $H_0^*$ & 1.4086 & 0.3345 & $H_1^*$ \\
 A-D  & 79.7641 & 49.0677 & $H_1^*$ & 33.5631 & 48.3546 & $H_0^*$ & 71.3135 & 52.2009 & $H_1^*$ \\
  \hline
\end{tabular}
\caption{Testing $H_0^*$ vs $H_1^*$ of pseudo observations of diverse financial instruments.}
\label{Test: SPQD-appl-3}
\end{table}

The null $H_0^*$ of $F_q^*=C^{\perp}$ for the pairs (CAD, GBP) and (JPY/USD, NASDAQ) is not rejected by any of the three tests, but the TOMD estimates are equal to 1.5488 and 1.4002, respectively, as seen from Tables~\ref{Tab:results 10per} and~\ref{MIX-Tab:results 10per}. These values suggest that the coordinates of the two aforementioned pairs may actually be fairly dependent.

\end{document}